# Regret in Treatment Choice when Welfare Varies with an Uncertain Event: The Prediction-Threshold Problem

Jeff Dominitz

Department of Economics and Ken Kennedy Institute, Rice University

Charles F. Manski

Department of Economics and Institute for Policy Research, Northwestern University

Abstract

We study maximum regret (MR) of binary treatment choice in a population with observed covariates x, when welfare varies with an uncertain binary event. We consider decision making with plug-in probabilistic predictions of the event and pre-specified decision thresholds, which we term the *prediction-threshold* problem. The optimal treatment for persons with covariate value x is B if the conditional probability P(y = 1|x) of a binary outcome y exceeds a particular x-specific threshold and is A otherwise. This structure is common in medical decision making and other contexts. Plug-in prediction uses data to estimate P(y|x) and acts as if the estimate is accurate. However, plug-in prediction is often performed with misspecified prediction models and conventional x-invariant thresholds. We use a combination of algebraic and computational analysis of limit and finite-sample MR to demonstrate how MR depends on the prediction model, the state space, and the thresholds used to choose treatments.

Acknowledgements: We have benefitted from the comments of Philip Marx, Jörg Stoye, and seminar participants at Northwestern University, as well as the excellent research assistance of Yeshaya Nussbaum.

## 1. Introduction

Abraham Wald's abstract concept of a statistical decision function (SDF) embraces all mappings from data into a chosen action. In practice, decision analysts commonly apply *inference-based* SDFs, which use sample data to infer the true state of nature and then choose an action that optimizes welfare as if the inference is correct. These SDFs have the form [data → inference → action]. See Manski (2021) for discussion.

An important subclass of inference-based SDFs are ones that use sample data to point-estimate a probability distribution P(y|x) of a welfare-relevant outcome y conditional on observed covariates x and use the estimate to make a decision that would be optimal if the estimate were accurate. Dominitz and Manski (2025) called these *prediction-based* SDFs. The approach is sometimes called *plug-in prediction*. We use the latter term here.

A restrictive feature of decisions made with plug-in predictions, which we maintain in this paper, is that they are generically deterministic functions of the data, assigning all persons with covariates x to the same treatment. Except in threshold cases where welfare calculated with plug-in predictions is equal across multiple treatments, plug-in treatment assignments cannot be fractional, randomly assigning observationally identical persons to different treatments. Nor can all observationally identical persons be randomly assigned to one treatment.

Econometricians, statisticians, and computer scientists have long studied plug-in prediction. However, they generally have not used statistical decision theory (SDT) to evaluate the performance of prediction models. Researchers have instead commonly viewed prediction as a self-contained inferential problem rather than as a task undertaken to inform decision making. A common practice has been to compute an estimate of a model that minimizes some measure of distance from predictions to observed outcomes. It has been particularly common to minimize the squared deviation between probabilistic predictions and outcomes. With binary outcomes and predictions, a common practice in machine learning (ML) of minimizing the overall misclassification rate may be represented as minimizing squared deviations between predicted and observed outcomes when predictions are constrained to take the value 0 or 1 (see Dominitz and Manski, 2025).

Frequentist econometricians and statisticians have long used confidence intervals and standard errors to measure the statistical precision of correctly specified prediction models, correct specification meaning that the actual conditional distribution P(y|x) is known to lie within the model space. When studying binary decisions between options A and B, they have used Neyman-Pearson hypothesis tests to judge whether B is better than A, the null hypothesis being that B is no better than A and the alternative being that it is better. These inferential methods are remote from SDT. Confidence intervals and standard errors play no role in the Wald theory. Considering binary decisions, Wald (1939) proposed SDT as a superior replacement for Neyman-Pearson testing.

In the 2000s, prediction is increasingly performed by computer scientists and other ML researchers who view prediction methods as computational algorithms and who do not use statistical theory to assess the algorithms. Instead, they perform *out-of-sample* tests of predictive accuracy using the common task framework (CTF) or K-fold cross-validation (CV). The CTF paradigm measures predictive accuracy in some pre-specified test data that may differ systematically from the training sample. They use CV to evaluate predictive performance on K alternative subsets of the realized training sample without considering how performance would vary across repeated draws of the training sample. We have previously critiqued these practices and recommended that ML research should use SDT to evaluate predictive algorithms. See Dominitz and Manski (2025).

Continuing research initiated in Manski (2023) and Dominitz and Manski (2025), this paper studies the maximum regret (MR) of plug-in prediction in a leading class of applied decision problems in which a planner chooses between two treatments for the members of a population with certain shared attributes that we leave implicit and observed covariates x that vary within the population. In these problems, the welfare function is monotone in the conditional probability P(y=1|x) of a binary outcome y. With such a welfare function, the optimal treatment for persons with covariate value x is B if

P(y=1|x) exceeds a particular x-specific threshold, say $t^*_x$, and is A otherwise. As in our past work, we use the classical Wald finite-sample framework. When we consider regret, it is relative to an ideal ex ante optimal decision given knowledge of P(y=1|x), not an institutionally constrained optimal such as in Kitagawa and Tetenov (2018).

This structure is common in medical decision making, where P(y=1|x) is the probability of occurrence of an illness. It appears that the first formal medical analyses, deriving thresholds that maximize expected patient welfare, were Pauker and Kassirer (1975, 1980) and Phelps and Mushlin (1988). Manski (2018) highlights applications to medical problems of choice between surveillance and aggressive treatment. For example, American guidelines for prophylactic treatment of breast cancer have used the Breast Cancer Risk Assessment (BCRA) Tool of the National Cancer Institute to predict probabilistic risk of illness and recommend aggressive treatment if the predicted probability is above a specified threshold.[1] In another example, the US Multi-Society Task Force on Colorectal Cancer reviewed studies of colonoscopy findings and colorectal cancer (CRC) risk to produce recommendations for post-colonoscopy follow-up and polyp surveillance (Gupta et al., 2020). Patel and Dominitz (2024) expressed the rationale for the recommendations as follows (p. ITC52): "Understanding CRC risk helps ensure that higher-risk patients (those with advanced or multiple polyps) undergo appropriate short-term follow-up but also ensures that lower-risk patients…are not exposed to the harms of excess colonoscopy."

The structure also arises in non-medical contexts such as criminal justice. Babii et al. (2026), Kleinberg et al. (2018), and Ludwig and Mullainathan (2021), for example, documented the use of risk predictions of future criminality by judges who make pre-trial release decisions. In proactive policing, the decision maker is a police officer who chooses whether to search an individual, and P(y=1|x) is the probability that the person has committed an offense, which may be determined through search. See Dominitz

[1] The BCRA Tool is publicly accessible online at https://bcrisktool.cancer.gov/. It uses the probabilistic predictions of a model initially estimated by Gail et al. (1989). A guideline issued by the American Society of Clinical Oncology has recommended consideration of a pharmacological intervention when the predicted probability of invasive cancer in the next five years is above 0.0166 (Visvanathan et al., 2009).

(2003), Knowles, Persico, and Todd (2001), and Manski and Nagin (2017).

Yet other applications occur in credit scoring and fraud detection. In credit scoring, the decision is whether to make a loan to an applicant. The event probability is whether the applicant will repay the loan. It is optimal to make a loan if the repayment probability exceeds some threshold. In fraud detection, it is optimal to take some action if specified information is communicated untruthfully and not to take the action otherwise. If truthfulness is not certain but can be accessed probabilistically, taking the action is considered optimal if the probability of truthfulness is less than a threshold.

As before, we study maximum regret because this measure of performance is conceptually appealing. By construction, an SDF with small MR is uniformly near-optimal across all possible populations generating training data and all populations of concern in decision making. Regret has a highly intuitive form in the treatment-choice problem that we study: the probability across samples that an SDF makes a sub-optimal decision times the magnitude of the welfare loss when a decision error occurs. The challenge to studying MR is analytical rather than conceptual. MR is algebraically complex except in certain simple settings. Computation commonly requires numerical methods to approximate MR.

As in our earlier work, we are concerned that the model used to estimate $P(y=1|x)$ may be misspecified, with true conditional probabilities being outside the model space. In practice, plug-in prediction has been performed with a wide variety of prediction models that commonly are misspecified. Medical research has long used parametric logit models. Econometricians have studied semiparametric linear-index models. Statisticians and econometricians have studied many versions of nonparametric regression models that assume $P(y=1|x)$ to vary smoothly with x. From the 1990s onward, it has become increasingly common to estimate “dimension-reducing models” when the covariate vector x has high dimension, these models making a “sparsity” assumption that $P(y=1|x)$ varies only with some low dimensional sub-vector of x. Recently, computer scientists have advocated use of deep neural network models, which replaced the assumption of dimensional sparsity with an alternative assumption of “compositional sparsity,” as discussed in Dominitz and Manski (2025). Among these prediction models, only nonparametric regression methods assuming local smoothness can credibly claim to be correctly specified in general. Application of these methods has been inhibited by concern with statistical imprecision, often described as the "curse of dimensionality."

The present research differs from our previous work in three primary ways, two substantive and one technical. The first substantive change is to expand the domain of the welfare function. We previously took the objective to be optimization of choice between two treatments for observationally identical persons, being those with a specified value of the covariates x. This is the relevant objective when, for example, a clinician wants to optimize treatment for a particular patient. In this paper, the objective is optimization of mean population welfare when treating a population with heterogeneous observable covariates. This may be the relevant objective when clinicians treat a group of patients or when a panel developing clinical practice guidelines recommends a treatment protocol for a broad population.

Expansion of the planning problem from treatment of observationally identical persons to treatment of an observably heterogeneous population does not complicate analysis of optimal treatment choice, as the optimal policy separates the population into observationally identical subgroups. However, it complicates regret analysis considerably when the conditional probabilities $P(y=1|x)$, $x \in X$ are not known and when population-wide data are used to estimate misspecified probability models. Then population MR does not decompose into the sum of x-specific MR values, making study of population MR complex.

The second substantive change is to expand the manner in which predictions are used to make treatment choices. Recall that, in the decision problems under study, the optimal treatment for persons with covariate value x is B if $P(y=1|x)$ exceeds some threshold $t^*_x$ and is A otherwise. We earlier assumed that the decision maker, henceforth called the planner, uses this threshold when applying plug-in prediction. The planner chooses B if the model estimate of $P(y=1|x)$ exceeds $t^*_x$ and A otherwise. Applications of plug-in prediction, including clinical practice guidelines, commonly use some conventional x-

invariant threshold. We are concerned that neither $t^*_x$ nor a conventional x-invariant threshold such as ½ need perform well from the perspective of MR when a misspecified plug-in predictor is used to estimate P(y=1|x). Rather than presume use of pre-specified thresholds, we study choice of thresholds to minimize MR. We henceforth refer to the planner's joint choice of a plug-in probabilistic prediction model and a threshold for treatment choice as the *prediction-threshold* problem.

The technical change is motivated by the substantive ones, which make analysis of MR more complex than in our earlier work. Study of MR using finite-sample data is highly challenging when the aim is to optimize a population-wide objective function and the prediction model may be misspecified. To simplify algebraic and numerical analysis, we begin with a limit setting in which the available data are generated by random sampling and sample size goes to infinity, enabling determination of the probability limit of the estimate.

Study of regret in the limit setting is instructive regarding finite-sample regret with large finite samples, when using estimates of predictor models that converge to deterministic limits. In that case, finite-sample regret converges to limit regret as sample size goes to infinity. To provide a sense of the adequacy of the limit approximation and evidence on the implications of statistical imprecision for threshold choice and treatment choice, we report numerical calculations of finite-sample regret.

The paper sheds new light on limit MR when plug-in prediction is performed with misspecified models and non-optimal decision thresholds. When a model is correctly specified, various estimation methods may provide consistent estimates of conditional event probabilities.[2] Then plug-in prediction yields zero limit MR when optimal thresholds are used for treatment choice, whatever the welfare function may be.

When a model is misspecified, the prediction-threshold problem is subtle. Estimates of event probabilities are generically inconsistent and limit MR commonly is positive, the specifics depending on the welfare function. The value of the limit estimate depends not only on the model but also on the estimation method. For example, the limit parameter estimate of an incorrectly specified binary logit model depends on whether one estimates the parameters by least squares, least absolute deviation, maximum likelihood, or another approach. Limit MR depends on both the limit estimate and on the thresholds used to choose treatments. Thresholds that are optimal with a correctly specified prediction model may be suboptimal with a misspecified model.

[2] A stream of research in decision and game theory beginning with Blackwell (1956) refers to consistent estimation of event probabilities as *calibration*. See, for example, Foster and Vohra (1998) and Foster and Hart (2018).

We use a combination of algebraic and computational analysis to study MR. We report informative algebraic analysis of limit MR when a planner applies certain elementary predictor models, estimation methods, and thresholds. However, algebraic analysis appears intractable in general. Hence, we turn to computation to study the behavior of various approaches that are often applied in practice.

Section 2 provides background, formalizing plug-in prediction for binary treatment choice as in Manski (2018). Section 3 presents our algebraic analysis of limit regret when prediction models are correct or misspecified. We first show that limit MR is zero using a correct model, whatever the welfare function may be. We then characterize the positive limit MR that occurs using elementary versions of two canonical misspecified prediction models. These models would be correct if the planner were to condition on subsets of the available covariate information, but they are misspecified conditional on all the covariates the planner observes.

Section 4 turns to computational study of limit and finite-sample MR of prediction-threshold pairs with a scalar covariate x that takes on four values. We compute MR in both the unrestricted state space and one restricted to weakly increasing functions of x, under four different specifications of known welfare. In addition to the correctly specified cell probability model, we include four misspecified models: marginal probability, linear and logit least squares, and a binary classifier. With the unrestricted state space, we find that limit MR with misspecified prediction models and/or suboptimal thresholds deviates greatly from the zero MR obtained with the correct prediction-threshold pair. With the monotonicity restriction on the state space, the limit MR of misspecified models is often close to zero with appropriately chosen thresholds.

The finite-sample computations produce two striking findings. First, the results are typically close to those for limit MR, suggesting that our analysis of limit MR in Section 3 provides a useful approximation to sample MR when sample size is only moderate. Second, finite-sample MR is sometimes smaller than limit MR when the prediction-threshold pair is suboptimal. The reason is that sampling imprecision can reduce the frequency of the errors that occur when limit predictions are on the wrong side of the threshold. This phenomenon may initially seem surprising. We explain in Section 4 why it is not anomalous.

The empirical problem motivating the analysis of this paper is incomplete knowledge of the conditional probability of a welfare-relevant event, with the available knowledge combining maintained assumptions and sample data. A further empirical problem that often arises in utilitarian treatment choice is incomplete knowledge of the x-specific welfare functions of members of the population. Maintained assumptions and sample data may be used to point-estimate these unknowns as well, thereby extending the scope for plug-in decision making. The mathematical structure of the treatment-choice problem studied in this paper is simple enough that much of our analysis can be applied with small modification to settings with incomplete knowledge of welfare functions. We explain in Section 5. Section 6 concludes.

**Related Literature.** The binary treatment choice problem studied in this paper arises in many applied contexts. Plug-in prediction is common. A diverse research literature spanning several fields has studied various approaches for sequential choice of a prediction model and decision thresholds. Our analysis differs in our use of maximum regret to evaluate the performance of strategies that jointly choose a possibly misspecified prediction model and decision thresholds. We think it important to appropriately summarize and critique the literature, focusing attention on approaches to measurement of prediction performance and to selection of decision thresholds. To understand our summary and critique requires the framework we develop in Sections 2 through 4. Hence, rather than discussing the prediction-threshold literature here, we do so in Supplemental Appendix A, which explains the respects in which we find the literature reasonable or wanting.

Before proceeding, we note that econometric research studying frequentist statistical decision theory has become increasingly common recently. Much of the focus has been on binary treatment choice with data from classical randomized trials or quasi-experiments, where the distribution of treatment response is point-identified and the sole concern is statistical imprecision. Contributions include Manski (2004, 2005, 2019), Hirano and Porter (2009, 2020), Stoye (2009), Manski and Tetenov (2016, 2019, 2021), Kitagawa and Tetenov (2018), and Mbakop and Tabord-Meehan (2021).

Attention is increasingly being given to treatment choice with data from trials with imperfect internal validity or from observational studies, where identification is the dominant concern with large sample size and joint attention to identification and statistical imprecision is essential when sample size is small. Contributions include Athey and Wager (2021), Manski (2000,2007,2021,2023), Montiel Olea, Qiu, and Stoye (2026), Stoye (2012), and Yata (2025).

It is important to recognize that the mathematical structure of the prediction-threshold problem we study here differs from other classes of binary treatment choice that econometricians have analyzed. For example, it differs substantially from the class of problems addressed by Montiel Olea, Qiu, and Stoye (2026) and Yata (2025), who consider problems in which welfare does not have the prediction-threshold structure and the sampling distribution of the data is normal with known variance.

## 2. Plug-In Prediction for Binary Treatment Choice

To study plug-in prediction, we need to formalize prediction and embed it in a decision problem. Considered in abstraction, the standard approach postulates a population characterized by a joint probability distribution P(y,x), where (y,x) takes values in some set Y×X. The conditional distribution P(y|x) probabilistically predicts y conditional on x. The state space S contains a set of feasible distributions $[P_s(y|x), x \in X]$,

$s \in S$. The unknown true distributions $[P(y|x), x \in X]$ are known to lie in this set.

We embed prediction in a decision problem by supposing that a planner wants to maximize welfare $W[\cdot; P(y|x), x \in X]$ over a choice set C. This is a statistical decision problem if the planner observes data $\psi$ drawn from a sample space $\Psi$ by some sampling process. Plug-in prediction uses the data to estimate $P(y|x)$ and acts as if the estimate is accurate. Let $\varphi_x(\cdot): \Psi \to S$ be the predictor function used to estimate $P(y|x)$. The chosen action solves the problem $\max_{c \in C} W[\cdot; \varphi_x(\psi), x \in X]$.

## 2.1. Binary Treatment Choice

We study a relatively simple yet subtle and illuminating version of this abstract decision problem, which arises in many applied contexts. A planner chooses between two treatments for each member of the population. For concreteness, we describe the planner as a clinician caring for a population of patients with observed covariates x. There are two care options, denoted A and B, for a particular disease. The welfare assumptions that we maintain are realistic when A is a noninvasive option and B an invasive one. As in Manski (2018) and our more recent work, we refer to A as surveillance and B as aggressive treatment.

We suppose that the clinician must choose a treatment without knowing a patient's illness status; y=1 if a patient is ill and y=0 if not. Observing x, the clinician can attempt to learn the conditional probability of illness, $p_x \equiv P(y=1|x)$. Medical research often proposes plug-in prediction, using sample data to estimate $p_x$ and acting as if the estimate is correct. This practice occurs in widely accepted clinical practice guidelines, as discussed in Section 1.

A version of this decision problem studied in Manski (2023) and Dominitz and Manski (2025) provides the starting point for our present study of plug-in prediction. We summarize this here and in Section 2.2. We generalize to the version of the present paper from Section 2.3 onward.

Let expected patient welfare with care option $c \in \{A,B\}$ have the known form $U_x(c,y)$; thus, expected welfare may vary with whether the disease occurs and with the patient covariates x. We assess patient welfare from the perspective of the clinician rather than from that of the patient. If the clinician is utilitarian and patients assess welfare conditional on x, $U_x(c,y)$ is also expected welfare from the perspective of patients.

The form of welfare function $U_x(\cdot,\cdot)$ depends on the context, but it is often reasonable to suppose that aggressive treatment is better if the disease occurs, and surveillance is better otherwise. That is,

(1a) $U_x(B,1) > U_x(A,1)$,

(1b) $U_x(A,0) > U_x(B,0)$.

We assume that these inequalities hold. We also assume that the chosen care option does not affect whether the disease occurs; hence, a patient's illness probability is simply $p_x$ rather than a function $p_x(c)$ of the care option. Treatment choice still matters because it affects the severity of illness and the patient's experience of side effects. Aggressive treatment is beneficial to the extent that it lessens the severity of illness, but harmful if it yields side effects that do not occur with surveillance.

2.1.1. Optimal Treatment Choice with Knowledge of $p_x$ and $U_x(\cdot,\cdot)$

Before considering plug-in prediction, suppose that the clinician knows $[p_x, U_x(\cdot,\cdot)]$ and maximizes expected welfare conditional on x. Then an optimal decision is

(2a) Choose A if $p_x \cdot U_x(A,1) + (1-p_x) \cdot U_x(A,0) \geq p_x \cdot U_x(B,1) + (1-p_x) \cdot U_x(B,0)$,

(2b) Choose B if $p_x \cdot U_x(B,1) + (1-p_x) \cdot U_x(B,0) \geq p_x \cdot U_x(A,1) + (1-p_x) \cdot U_x(A,0)$.

The decision yields optimal expected patient welfare

(3) $\max\,[p_x \cdot U_x(A,1) + (1-p_x) \cdot U_x(A,0),\; p_x \cdot U_x(B,1) + (1-p_x) \cdot U_x(B,0)]$.

Both A and B are optimal if these treatments yield equal expected welfare.

The optimal decision is easy to characterize when inequalities (1a) and (1b) hold. Let $p_x^{\#}$ denote the threshold value of $p_x$ that makes options A and B have the same expected welfare. This value is

$$(4) \qquad p_x^{\#} = \frac{U_x(A,0) - U_x(B,0)}{[U_x(A,0) - U_x(B,0)] + [U_x(B,1) - U_x(A,1)]}$$

Option A is optimal if $p_x \leq p_x^{\#}$ and B if $p_x \geq p_x^{\#}$. Thus, optimal treatment choice does not require exact knowledge of $p_x$. It only requires knowing whether $p_x$ is larger or smaller than the threshold $p_x^{\#}$. So the prediction-threshold problem has a simple optimal solution. To the best of our knowledge, Pauker and Kassirer (1975, 1980) gave the first derivation of threshold (4), followed by Phelps and Mushlin (1988). In the literature on ML research, Elkan (2001) derived the same threshold in a paper that uses the terminology "cost-sensitive learning."

2.1.2. Aggressive Treatment Neutralizes Disease

An instructive simplifying case occurs when aggressive treatment neutralizes disease, in the sense that $U_x(B,0) = U_x(B,1)$. For example, aggressive treatment might be surgery to remove a localized tumor that may (y=1) or may not (y=0) be malignant. Suppose that surgery always eliminates cancer when present. Then surgery neutralizes disease. Being invasive and costly, performance of surgery has a negative side effect on welfare that is the same regardless of whether cancer is present.

Let $U_{xB}$ denote welfare with aggressive treatment. Then (1) – (4) reduce to

$$(5) \quad U_x(A,0) > U_{xB} > U_x(A,1).$$

$$(6a) \quad \text{Choose A if } p_x \cdot U_x(A,1) + (1-p_x) \cdot U_x(A,0) \ \geq \ U_{xB},$$

(6b) Choose B if $U_{xB} \geq p_x \cdot U_x(A,1) + (1-p_x) \cdot U_x(A,0)$.

(7) $\max\,[p_x \cdot U_x(A,1) + (1-p_x) \cdot U_x(A,0),\ U_{xB}]$.

(8) $$p_x^{\#} = \frac{U_x(A,0) - U_{xB}}{U_x(A,0) - U_x(A,1)}.$$

Further simplification occurs when one normalizes the location and scale of welfare by setting $U_x(A,0) = 1$ and $U_x(A,1) = 0$. This normalization is used routinely in health economics research on *quality of life*, where the welfare value 1 is interpreted as perfect health and the value 0 as health so poor that a person would be indifferent between being alive and dead; see Weinstein, Torrance, and McGuire (2009). Then (5) − (8) become

(9) $1 > U_{xB} > 0$.

(10a) Choose A if $1-p_x \geq U_{xB}$,

(10b) Choose B if $U_{xB} \geq 1-p_x$.

(11) $\max\,(1-p_x, U_{xB})$.

(12) $p_x^{\#} = 1-U_{xB}$.

To illuminate basic issues, we henceforth assume that aggressive treatment neutralizes disease and normalize the welfare of surveillance as above. The methodological approach that we will develop to study the MR of plug-in prediction may in principle be used to study more general decision problems than binary treatment choice, with or without such simplifications.

## 2.2. Maximum x-Specific Regret with Plug-In Prediction

In our earlier work, we considered a clinician treating patients with covariates x. These patients are observationally identical. The clinician knows $U_x(\cdot,\cdot)$, but not $p_x$. The state space $S_x$ lists all feasible values of $p_x$, denoted $p_{sx} \equiv P_s(y=1|x)$, $s \in S_x$.

A clinician who does not know whether $p_x$ is smaller or larger than the threshold $p_x^{\#} = 1-U_{xB}$ faces a problem of decision making under ambiguity. Suppose the clinician knows that $p_{xm} < 1-U_{xB} < p_{xM}$, where $p_{xm} \equiv \min_{s \in S} p_{sx}$ and $p_{xM} \equiv \max_{s \in S} p_{sx}$. For instance, it is well known that the Law of Total Probability implies that knowledge of the marginal probabilities $P(y=1)$ and $P(x)$, $x \in X$, generically yields an informative identification region $[p_{xm}, p_{xM}]$ for $p_x$; see Supplemental Appendix B for details. Still, the clinician lacks the information needed to maximize expected welfare conditional on x.

Manski (2018) derived the minimax regret treatment using the above information alone—that is, without sample data—and requiring that treatment assignment be singleton. The proof is simple. Choosing A is an error if $p_x > 1-U_{xB}$, yielding regret $p_x - (1-U_{xB})$. MR is $p_{xM} - (1-U_{xB})$, occurring when $p_x$ attains its maximum feasible value. Symmetrically, choosing B is an error if $p_x < 1-U_{xB}$, yielding regret $(1-U_{xB}) - p_x$. MR is $(1-U_{xB}) - p_{xm}$, occurring when $p_x$ attains its minimum value. Hence, the treatment choice that minimizes MR is B if $(1-U_{xB}) - p_{xm} < p_{xM} - (1-U_{xB})$, equivalently if $(p_{xm}+p_{xM})/2 > 1-U_{xB}$. Symmetrically, A minimizes MR if $(p_{xm}+p_{xM})/2 < 1-U_{xB}$. Therefore, minimax regret (MMR) equals $\min\{p_{xM}-(1-U_{xB}), (1-U_{xB})-p_{xm}\}$. Restricting treatment choice to singleton rules is consequential. Without this restriction, Manski (2009) showed that the MMR rule is fractional whenever both treatments are undominated.

It has been common in applied prediction to place no a priori restrictions on event probabilities, so $p_{xm} = 0$ and $p_{xM} = 1$. In this case with no knowledge of $p_x$, B minimizes MR if $\frac{1}{2} > 1-U_{xB}$, A if $\frac{1}{2} < 1-U_{xB}$, and MMR simplifies to $\min(U_{xB}, 1-U_{xB})$.

Plug-in prediction uses data to estimate $p_x$ and acts as if the estimate is correct. Let $Q_s$ be a sampling distribution generating data $\psi \in \Psi$. Let $\varphi_x(\psi)$ be a point estimate of $p_x$. In practice, researchers typically use data drawn from the entire population to estimate prediction models, even if the objective is only to learn the outcome probability at a particular value of x. Hence, we use the general state space notation S from now on, with s an element of S, rather than the x-specific notation $S_x$.

Plug-in prediction maximizes expected welfare acting as if $p_x = \varphi_x(\psi)$, in which case the as-if optimal threshold is $p_x^{\#} = 1 - U_{xB}$. Until now, with correct prediction, it was not necessary to specify the chosen treatment when equality occurs. However, with possibly incorrect plug-in predictions, this choice becomes consequential in circumstances where equality may hold. In this paper, we assume that treatment A is chosen when equality occurs, which is consistent with the traditional practice in hypothesis testing where treatment A is the null hypothesis and B is the alternative. Then the treatment choice is A if $\varphi_x(\psi) \leq 1 - U_{xB}$ and is B if $\varphi_x(\psi) > 1 - U_{xB}$. We seek to determine conditions under which treatment choice with plug-in prediction reduces MR below that of treatment choice with no knowledge of $p_x$ and, if so, by how much.

Consider the regret of treatment choice acting as if $p_x = \varphi_x(\psi)$. Let $e[p_{sx}, \varphi_x(\psi), U_{xB}]$ denote the occurrence of an ex ante decision error in state s when $\varphi_x(\psi)$ is used to choose treatment. That is, $e[p_{sx}, \varphi_x(\psi), U_{xB}] = 1$ when $p_{sx}$ and $\varphi_x(\psi)$ yield different treatments, while $e[p_{sx}, \varphi_x(\psi), U_{xB}] = 0$ when $p_{sx}$ and $\varphi_x(\psi)$ yield the same treatment. It is important to recognize that the occurrence of errors does not necessarily increase with the distance between $\varphi_x(\psi)$ and $p_{sx}$. An error occurs only if the magnitude and sign of the difference between these quantities places them on opposite sides of the decision threshold.

Regret using estimate $\varphi_x(\psi)$ and threshold $p_x^{\#} = 1 - U_{xB}$ is

$$
\begin{aligned}
(13)\quad R_{sx}[\varphi_x(\psi)] &= \max(1 - p_{sx}, U_{xB}) - (1 - p_{sx}) \cdot 1[1 - \varphi_x(\psi) \geq U_{xB}] - U_{xB} \cdot 1[U_{xB} > 1 - \varphi_x(\psi)] \\
&= |(1 - U_{xB}) - p_{sx}| \cdot 1[p_{sx} \leq 1 - U_{xB} < \varphi_x(\psi) \text{ or } \varphi_x(\psi) < 1 - U_{xB} \leq p_{sx}] \\
&= |(1 - U_{xB}) - p_{sx}| \cdot e[p_{sx}, \varphi_x(\psi), U_{xB}].
\end{aligned}
$$

Regret across samples is the mean of this quantity with respect to the sampling distribution, namely

(14) $E_s\{R_{sx}[\varphi_x(\psi)]\} = |(1-U_{xB})-p_{sx}|\cdot Q_s\{e[p_{sx}, \varphi_x(\psi), U_{xB}] = 1\}$.

Here $Q_s\{e[p_{sx}, \varphi_x(\psi), U_{xB}] = 1\}$ is the probability across samples that an error occurs. The non-negative quantity $|(1-U_{xB}) - p_{sx}|$ is the magnitude of the loss in welfare when an error occurs. Maximum regret across the state space is $\max_{s\in S} E_s\{R_{sx}[\varphi_x(\psi)]\}$.

Observe that expected regret (14) is determined not only by the frequency of treatment errors but also by the magnitude of the welfare losses when errors occur. In contrast, research on statistical hypothesis testing has been concerned only with the frequency of errors, not with the magnitudes of the associated welfare losses (Neyman and Pearson, 1928, 1933). Research on empirical risk minimization aims to estimate a predictive model that minimizes the sample misclassification rate averaged across persons with different covariate values; that is, the empirical frequency of errors (Vapnik, 1999, 2000). Thus, these methodologies ignore the magnitudes of the welfare losses when errors occur, which generally leads to suboptimal decision making with respect to maximum regret.

## 2.3. Choice of Predictor and Thresholds for Decision Making in a Population

The analysis in this paper modifies the above setting in two ways, in order to emulate the environment that has typically been studied in research on plug-in prediction. First, we consider decision making in a population whose members have heterogenous values of x taking values in a finite set X, rather than in a sub-population with a specified covariate value. We consider the covariate distribution P(x) to be known, with $P(x) > 0$ for all $x\in X$. In this setting, with an unrestricted state space and no data, minimax regret is $\sum_{x\in X} P(x) \min\{U_{xB}, 1-U_{xB})$. This holds because, with no cross-covariate restrictions, minimization of population MR decomposes to summation of covariate-specific MR. See Section 2.2 above for covariate-specific MR and Manski (2024, ch. 7) for the decomposition result.

Second, we do not presuppose that the planner necessarily uses $1-U_{xB}$ as the probability threshold separating choice of treatment A and B. This threshold would be optimal if event probabilities were known, but it may not perform well from the MR perspective when a plug-in predictor is used to estimate $P(y=1|x)$. Rather than presume use of pre-specified thresholds, we study choice of thresholds to minimize MR. Thus, we suppose that the planner chooses a vector of x-specific threshold values $t_x \in [0,1]$, $x \in X$. For each $x \in X$, the planner chooses treatment A if $\varphi_x(\psi) \leq t_x$ and chooses B if $\varphi_x(\psi) > t_x$. Joint choice of a predictor function and thresholds $[\varphi_x(\cdot), t_x, x \in X]$ constitutes the prediction-threshold problem.

Extending the notation for error probabilities and regret to encompass choice of thresholds, population regret with the predictor function and thresholds $[\varphi_x(\psi), t_x; x \in X]$ is

$$(15) \quad \sum_{x \in X} P(x)\, E_s\{R_{sx}[\varphi_x(\psi), t_x]\} = \sum_{x \in X} P(x)\, |(1-U_{xB})-p_{sx})| \cdot Q_s\{e[p_{sx}, \varphi_x(\psi), t_x, U_{xB}] = 1\}.$$

Let $\Phi$ denote a set of predictor functions that the planner contemplates using. A predictor function and thresholds minimize maximum regret if they solve the problem

$$(16) \quad \min_{\varphi \in \Phi,\, t_x \in [0,1],\, x \in X} \; \max_{s \in S} \; \sum_{x \in X} P(x)\, |(1-U_{xB}) - p_{sx}| \cdot Q_s\{e[p_{sx}, \varphi_x(\psi), t_x, U_{xB}] = 1\}.$$

This formalization of the decision problem adheres to the frequentist nature of Wald's statistical decision theory. The planner commits to a decision criterion ex ante, before the data are observed, and evaluates the criterion by its performance across repeated samples.

In the common applied practice where researchers use population-wide data to estimate prediction models, problem (16) is considerably more complex than minimization of x-specific MR. In each state of nature s, population regret is a weighted average of x-specific regret across the covariate space X, with the covariate distribution giving the weights. Maximization of population regret over S requires joint consideration of all x-specific regret values, not each element of X in isolation. Minimization over predictor models and thresholds in the outer minimization step of the MMR problem again requires joint consideration of all elements of X, not each element in isolation.

## 3. Limit Regret with Random Sampling of Events

It is generally infeasible to solve problem (16) analytically. The problem is computationally challenging as well. The error probabilities $Q_s\{e[p_{sx}, \varphi_x(\psi), t_x, U_{xB}] = 1\}$ generally do not have explicit forms. In many cases of applied interest, Monte Carlo simulation can be used to easily approximate the error probability for any specified covariate value, threshold, predictor model, and state. The difficulty is that it is not possible to perform Monte Carlo analysis repeatedly across the uncountably many elements of the product set $[x \in X, \{t_x \in [0,1], x \in X\}, \varphi \in \Phi, s \in S]$. It may be tractable to approximately solve (16) by restricting attention to a finite grid of (covariate, threshold, predictor model, state) values. We illustrate in Section 4.

To lessen the analytical and computational burden, in this section we suppose that data are drawn at random from P(y,x) and we study the limit of maximum regret as sample size goes to infinity. Let $\psi = (y_i,x_i)$, i = 1 . . . N be a random sample of size N. Let $\Phi$ be a set of predictor functions that converge to limit predictors as sample size goes to infinity. Thus, the probability limit of the empirical distribution of $\psi$ in state of nature s is the population distribution $P_s$ and that of $\varphi_x(\psi)$ is some deterministic conditional probability $\varphi_{sx}$. Given a prediction-threshold pair, the x-specific limit error in state s is 1 if $(p_{sx} \le 1 - U_{xB}, \varphi_{sx} > t_x)$ or if $(1 - U_{xB} < p_{sx}, \varphi_{sx} \le t_x)$. It is 0 otherwise. Hence, the limit version of problem (16) is

$$(17) \quad \min_{\varphi \in \Phi,\ t_x \in [0,1],\ x \in X} \left\{ \max_{s \in S} \sum_{x \in X} P(x)|(1-U_{xB}) - p_{sx}| \cdot 1[(p_{sx} \le 1-U_{xB}, \varphi_{sx} > t_x) \text{ or } (1-U_{xB} < p_{sx}, \varphi_{sx} \le t_x)] \right\}.$$

We henceforth study problem (17).[3]

In what follows, Section 3.1 examines threshold selection when the prediction model is correctly specified, meaning that $(\varphi_{sx}, x \in X) = (p_{sx}, x \in X)$ for all $s \in S$. This setting yields the optimistic result that zero maximum regret is achievable, whatever the welfare function may be. However, correct specification of the predictor model is rarely credible except when researchers use nonparametric regression methods to predict outcomes, in which case weak regularity conditions imply that $\varphi_x(\psi)$ consistently estimates $p_{sx}$. Our main concern is the use of plug-in prediction with misspecified prediction models.

Sections 3.2 and 3.3 report algebraic analysis of the positive MR that results using elementary versions of two canonical misspecified prediction models. These models would be correctly specified if the planner were to condition on subsets of the available covariate information, but they are misspecified conditional on all the covariates the planner observes. The findings are instructive.

We find it intractable to algebraically analyze more complex types of misspecified models that are commonly used in practice. Nevertheless, numerical computation can be revealing. Section 4 presents numerical findings on maximum limit and finite-sample regret using familiar parametric prediction models estimated by least squares, in combination with alternative thresholds for treatment choice.

A prevalent practice in nonparametric probabilistic prediction with a finite covariate space has been to place no a priori restrictions on conditional event probabilities. Thus,

---

[3] Problem (17) is the probability limit of (16) given certain regularity conditions. For a specified predictor model and threshold, compare finite-sample x-specific regret (14) with the limit version replacing $\varphi_x(\psi)$ by $\varphi_{sx}$. Defining $\varepsilon_{sx} \equiv |(1-U_{xB})-p_{sx}|$, the difference between finite-sample and limit regret is

$$\varepsilon_{sx}\,[Q_s\{e[p_{sx}, \varphi_x(\psi), U_{xB}]=1\} - e[p_{sx}, \varphi_{sx}, U_{xB}]].$$

When $\varepsilon_{sx}=0$, this difference is zero regardless of the bracketed difference in finite-sample and limit error probabilities. When $\varepsilon_{sx}>0$, $e[p_{sx}, \varphi_x(\psi), U_{xB}] = e[p_{sx}, \varphi_{sx}, U_{xB}]$ if $\varphi_x(\psi)$ and $\varphi_{sx}$ are on the same side of the threshold $t_x$. Hence, the regret difference must go to zero as $N\to\infty$ provided that $\varphi_{sx}\neq t_x$. If this regularity condition holds for all elements of X and if the convergence of $\varphi_x(\psi)$ to $\varphi_{sx}$ is uniform in S and $\Phi$, it follows that (16) converges to (17). The relationship of problems (16) and (17) may be more delicate if some limit predictions exactly equal thresholds or if the convergence of finite-sample predictors to their limits is not sufficiently uniform.

the state space S is the unit hypercube $[0,1]^{|X|}$. We specify S this way throughout the remainder of Section 3, but not always in Section 4.

3.1. Limit Threshold Choice with Correct Predictions

Problem (17) is simple if $\varphi_{sx} = p_{sx}$ is a feasible predictor for all (s, x) and if the state space places no cross-covariate restrictions on $p_{sx}$, $x\in X$. Supposing that $\varphi_{sx} = p_{sx}$ to be feasible is a realistic approximation if sample size is sufficiently large and the cardinality of X is sufficiently small that accurate nonparametric estimation of $p_{sx}$ is feasible. The most obvious case where the state space makes no cross-covariate restriction is the one routinely assumed in nonparametric estimation with a finite covariate space, in which the researcher considers possible all vectors $(p_x, x\in X)$ in the unit hypercube.

3.1.1. Optimal Thresholds

When the state space is unrestricted and $\varphi_{sx} = p_{sx}$, (17) decomposes into a P(x)-weighted average of x-specific MMR problems, namely

$$(18)\quad \sum_{x\in X} P(x) \min_{t_x\in[0,1]} \max_{p_{sx}\in[0,1]} |(1-U_{xB})-p_{sx}|\cdot 1[t_x<p_{sx}\le 1-U_{xB} \text{ or } 1-U_{xB}<p_{sx}\le t_x].$$

For each value of $t_x$, the inner maximization occurs when $p_{sx} = t_x$. This is the value of $p_{sx}$ that maximizes the distance between $p_{sx}$ and $(1\text{-}U_{xB})$, subject to the constraint that either $p_{sx}\in(t_x, 1-U_{xB}]$ or $p_{sx}\in(1-U_{xB}, t_x]$. Hence, maximum x-specific regret is

$$(19)\quad \max_{p_{sx}\in[0,1]} |(1-U_{xB})-p_{sx}|\cdot 1[t_x<p_{sx}\le 1-U_{xB} \text{ or } 1-U_{xB}<p_{sx}\le t] = |(1-U_{xB})-t_x|.$$

Maximum population regret with thresholds $(t_x, x\in X)$ is $\sum_{x\in X} P(x)|(1-U_{xB})-t_x|$. Maximum population regret is minimized by setting $t_x = 1-U_{xB}$, in which case maximum regret is zero. Recall that this threshold choice is the optimal threshold in (12) with knowledge of $p_x$.

The above finding of zero limit MR with a correct prediction model and optimal thresholds holds whatever the welfare function may be; that is, for all values of ($U_{xB}$, $x \in X$). A salutary implication is that availability of a correct prediction model enables costless separation of the tasks of prediction and treatment choice. Wald's statistical decision theory assumes that the same individual observes data and makes decisions. In practice, however, it is common for a researcher to estimate a prediction model, not knowing the welfare function that a separate decision maker will use to choose treatments. The fact that a correct prediction model achieves zero limit regret for all welfare functions implies optimality of a sequential process in which the researcher communicates the correct prediction model to the decision maker, who then chooses decision thresholds.

We caution that a sequential process of prediction followed by threshold selection generally does not optimize MR when the available prediction models are misspecified. We demonstrate in Section 4 that joint choice of predictions and thresholds sometimes achieves smaller MR.

3.1.2. Optimization with x-Invariant Thresholds

It has been conventional in the medical literature to select an x-invariant threshold t, setting $t_x = t$ for all x. In this case, (17) becomes

$$(17') \quad \min_{\varphi \in \Phi,\ t \in [0, 1],\ x \in X} \left\{ \max_{s \in S} \sum_{x \in X} P(x)\, |(1-U_{xB})-p_{sx}| \cdot 1[(p_{sx} \leq 1-U_{xB},\ \varphi_{sx} > t) \text{ or } (1-U_{xB} < p_{sx},\ \varphi_{sx} \leq t)] \right\}.$$

The predictor $\varphi_{sx} = p_{sx}$ has maximum population regret $\sum_{x \in X} P(x)|(1-U_{xB})-t|$, which solves the MMR problem if $U_{xB} = 1-t$ for all x, but not otherwise. The MMR x-invariant threshold with $\varphi_{sx} = p_{sx}$ solves

$$(20) \qquad \min_{t \in [0,1]} \sum_{x \in X} P(x)|(1-U_{xB})-t|.$$

Maximum regret is nonzero unless $U_{xB}$ happens to be x-invariant.

If the optimal x-specific thresholds $(1-U_{xB})$ are sufficiently disparate, the solution to (20) may exceed MR with no data and an unrestricted state space, discussed in Section 2.3, where MR = $\sum_{x \in X} P(x) \min\{U_{xB}, 1-U_{xB}\}$. Consider, for example, binary x taking the values 0 and 1, with $P(x=0) = P(x=1) = ½$ and $U_{0B} = 1-U_{1B} > 0.5$. Then MR with no data is $1-U_{0B}$. By contrast, (20) is solved by setting t to be any point between the optimal x-specific thresholds. In this case, t=½ is a solution and limit MR = $U_{0B}-½$, which exceeds MR with no data if $U_{0B}>0.75$. Thus, constraining the planner to ignore the plug-in prediction with data may result in smaller MR than constraining the decision threshold with plug-in prediction to be x-invariant. This result is not surprising. Using the correction prediction model with a suboptimal threshold may be inferior to the MMR singleton rule with no knowledge of P(y|x).

Given a correct prediction model, this deficiency can be overcome by using the limit predictor $\varphi_{sx} = p_{sx}-[(1-U_{xB})-t]$ rather than $\varphi_{sx} = p_{sx}$. Then maximum regret is zero for any $t \in [0,1]$. Thus, if the correct conditional probabilities $p_{sx}$ are known, the apparent deficiency of using an x-invariant threshold may effectively be undone by modifying the predictor appropriately.

### 3.2. Using the Marginal Event Probability to Predict Conditional Probabilities with a Binary Covariate

The general MMR problem (17) is too complex to permit effective algebraic analysis, but some special cases yield instructive findings. A canonical type of misspecified prediction arises when prediction is performed conditional on a many-to-one function of x, say w(x). Then prediction at most reveals P[y=1|w(x)], not the more informative probabilities P(y=1|x). This section examines the elementary but non-trivial version of this setting in which x is a binary covariate, w(x) is a constant function, and

the predictor in state s is the correct marginal event probability $p_s \equiv P_s(y=1)$. In this case, (17) becomes

$$(21) \quad \min_{t_x \in [0,\,1],\; x \in X} \left\{ \max_{s \in S} \sum_{x \in X} P(x)\left|(1-U_{xB})-p_{sx}\right| \cdot 1\left[(t_x<p_s,\; p_{sx}\leq 1-U_{xB}) \text{ or } (p_s \leq t_x,\; 1-U_{xB}<p_{sx})\right] \right\}.$$

Note that feasible states of nature satisfy the adding up condition: $\sum_{x \in X} P(x) \cdot p_{sx} = p_s$.

Analysis of (21) is complex in general, but it simplifies if x is binary, taking the values 0 and 1. Then $\varphi_{sx} = p_s = P(x=0)p_{s0} + P(x=1)p_{s1}$. We further simplify by assuming that decisions are made with the x-invariant threshold $\tilde{t} = P(x=0)(1-U_{0B}) + P(x=1)(1-U_{1B})$, which averages the optimal x-specific thresholds with respect to the covariate distribution. We find that the form of MR is subtle.

We show in Supplemental Appendix B that the optimal treatment is assigned for at least one of the two values of x. Maximum population regret is determined by the values of P(x) and $U_{xB}$, taking one of four values: $P(x=0)(1-U_{0B})$, $P(x=1)(1-U_{1B})$, $P(x=0)U_{0B}$, or $P(x=1)U_{1B}$. Table 1 gives the conditions under which MR takes each value.

Table 1: Maximum Regret for Binary x with Prediction by the Marginal Probability and x-Invariant Threshold

| | $P(x=1) \leq \tilde{t}$ | $P(x = 1) > \tilde{t}$ |
|---|---|---|
| $P(x=0) \leq \tilde{t}$ | $\max\{P(x=0)U_{0B},\ P(x=1)U_{1B}\}$ | $P(x = 0)\max\{U_{0B},\ (1-U_{0B})\}$ |
| $P(x=0) > \tilde{t}$ | $P(x=1)\max\{U_{1B},\ (1-U_{1B})\}$ | $\max\{P(x=0)(1-U_{0B}),\ P(x=1)(1-U_{1B})\}$ |

Note: $\tilde{t} = P(x=0)(1-U_{0B}) + P(x=1)(1-U_{1B})$.

An instructive finding in the derivation of Table 1 concerns the values that the marginal and conditional probabilities take where regret is maximized. In particular, $p_s$ takes the threshold value $\tilde{t}$, analogous to the case with correct predictions and optimal thresholds where regret is maximized with $p_{sx} = t_x$. Here, regret is maximized where $p_{s0}$ and $p_{s1}$ take opposite extreme values as determined by the Law of Total Probability with $p_s = \tilde{t}$. Supplemental Appendix B shows that when one of the values for $p_{sx}$ is in the open interval (0,1), the other is at an end point 0 or 1.

Observe that MR is determined not only by the utilities ($U_{0B}$,$U_{1B}$) but also by the covariate distribution P(x). This distribution affects MR in two distinct ways. One is that, given any value of the marginal probability $P_s(y=1)$, P(x) determines the frequencies with which different treatment errors are made. The other is that $P_s(y=1)$ is a weighted average of $P_s(y=1|x=0)$ and $P_s(y=1|x=1)$, with weights P(x=0) and P(x=1). The entries in Table 1 reflect both considerations.

Also observe that, regardless of the threshold, MR is zero in the polar cases where P(x) is degenerate, all persons having the same covariate value. In these cases, the marginal probability equals the cell probability for the dominant x value. The marginal probability is uninformative about the other x value, but that value receives zero weight in the MR calculation.

### 3.2.1. Applications to Surveillance versus Removal of Local Tumors

In Section 2.1.2, we noted that the binary treatment choice problem studied in this paper approximates well the clinical decision whether to remove a localized tumor that may (y=1) or may not (y=0) be malignant. It is often reasonable to assume that surgery neutralizes disease and that performance of surgery has a negative side effect on welfare that is the same regardless of whether cancer is present.

It is common for articles reporting medical research to report a marginal probability of malignancy P(y=1) across a population of interest, but not to report probabilities P(y=1|x) conditional on observed patient covariates that vary within this population. Then Table 1 gives the MR that results if a clinician who observes a binary covariate x uses a reported marginal probability to predict malignancy and the x-invariant threshold $\tilde{t}$ to choose between surveillance and surgery.

### 3.3. Using a K-Dimensional Parametric Model to Correctly Predict K Conditional Probabilities

Another canonical type of misspecified prediction arises when prediction is performed conditional on a subset of the x values that occur in the population, say $X_0 \subset X$. Then prediction at most reveals $P(y=1|x)$, $x \in X_0$. This section examines the elementary but non-trivial version of this setting in which the predictions in $X_0$ are correct.

Let $K \geq 1$ be the cardinality of $X_0$. Let $F(\cdot,\cdot)$: $X \times B \rightarrow [0,1]$ be an invertible K-dimensional parametric predictor model. The parameter space B is a set in $R^K$. Let $(x_k, k = 1, \ldots, K)$ denote the first K elements of X. Invertibility means that, for all $\{\alpha_k, \in [0,1], k = 1, \ldots, K\}$, there exists a $b \in B$ such that the equations $F(x_k,b) = \alpha_k$, $k = 1, \ldots, K$ have a solution in B.

In state of nature s, using the model to predict event probabilities implies that there exists a $b_s \in B$ solving the equations

$$\text{(22)} \quad \varphi_{sx} = F(x_k,b_s) = p_{sx},\ x = x_1, \ldots, x_K.$$

However, $\varphi_{sx} = F(x_k,b_s)$ need not equal $p_{sx}$ for $k > K$. This is so because $b_s$ is determined entirely by $p_{sx}$, $x = x_1, \ldots, x_K$. Thus, the model is generically misspecified.

For any specified thresholds $(t_x, x \in X)$, the maximum regret of the limiting parametric predictor is

$$\text{(23)} \quad MR = \max_{s \in S} \sum_{x \in X} P(x)|(1-U_{xB})-p_{sx}|\cdot 1[(p_{sx} \leq 1-U_{xB},\ F(x,b_s)>t_x) \text{ or } (1-U_{xB} < p_{sx},\ F(x,b_s) \leq t_x)].$$

This expression simplifies if, for $(x = x_1, \ldots x_K)$, one uses the thresholds $t_x = 1-U_{xB}$ that were shown in Section 3.1.1 to be optimal with correct predictions. Given that $F(x,b_s) = p_{sx}$ for these values of x,

$$1[(p_{sx} \le 1-U_{xB},\ F(x,b_s)>t_x) \text{ or } (1-U_{xB} < p_{sx},\ F(x,b_s) \le t_x)] = 0$$

for all $s \in S$. Hence, the regret terms for these values of x in (23) equal zero for all s.

Now consider $x_k$, $k > K$. The solution $b_s$ to the prediction problem is a function only of ($p_{sx}$, $x = x_k$, $k = 1, \ldots, K$), not of ($p_{sx}$, $x = x_k$, $k > K$). Suppose that the state space does not constrain the latter conditional probabilities; that is, the feasible states of nature include all $p_{sx} \in [0,1]$, $x = x_k$, $k > K$. Then, for all thresholds ($t_x$, $x = x_k$, $k > K$), there exists feasible values of $p_{sx}$, $x = x_k$, $k > K$ such that

$$1[(p_{sx} \le 1-U_{xB},\ F(x,b_s)>t_x) \text{ or } (1-U_{xB} < p_{sx},\ F(x,b_s) \le t_x)] = 1.$$

When this occurs, the regret term for $x_k$, $k > K$, in state s is $P(x)|(1-U_{xB})-p_{sx}|$. If $F(x,b_s)>t_x$, this expression is maximized when $p_{sx} = 0$, in which case its value is $P(x)(1-U_{xB})$. If $F(x,b_s) \le t_x$, the expression is maximized when $p_{sx} = 1$, in which case its value is $P(x)U_{xB}$. If $U_{xB} = ½$, the maximum value $½P(x)$ occurs when $p_{sx}$ equals 0 or 1.

The above derivation implies that maximum population regret (23) reduces to

$$(24)\quad MR = \max_{s \in S} \sum_{x_k, k>K} P(x)\{(1-U_{xB})\cdot 1[F(x,b_s)>t_x] + U_{xB}\cdot 1[F(x,b_s) \le t_x]\}.$$

This expression is still a bit complex, but it implies a simple informative bound on maximum regret. For each value of x in the above summation, the regret term is either $P(x)(1-U_{xB})$ or $P(x)U_{xB}$, mirroring the findings for the marginal event probability predictor reported in Table 1. Hence, for all K-dimensional parametric predictors and all thresholds ($t_x$, $x = x_k$, $k > K$), MR is contained in the bound

$$(25)\quad \sum_{x_k, k>K} P(x) \min(1-U_{xB}, U_{xB}) \le MR \le \sum_{x_k, k>K} P(x)\max(1-U_{xB}, U_{xB}).$$

As was the case with the marginal probability predictor, the present bound is determined by a multiplicative interaction of the utilities and the covariate distribution.

This bound further simplifies if $U_{xB}$, ($x=x_k$, $k>K$) does not vary with x. Then (25) reduces to

$$(26) \quad \min(1-U_B,U_B) \sum_{x_k,k>K} P(x) \;\leq\; MR \;\leq\; \max(1-U_B,U_B) \sum_{x_k,k>K} P(x),$$

where $U_B$ is the x-invariant value of welfare.

Yet further simplification occurs if $U_B = ½$. Then $\min(1-U_B,U_B) = \max(1-U_B,U_B) = ½$. Hence, the bound (26) reduces to the unique value

$$(27) \quad MR = ½ \sum_{x_k,k>K} P(x).$$

Observe that the covariate distribution matters. Bounds (26) and (27) are minimized if $x_1, \ldots x_K$ are chosen to be the K values of x with the highest covariate probabilities P(x). This selection of the K values of x minimizes the summation $\sum_{x_{k,k>K}} P(x)$.

Another striking algebraic finding emerges without restricting $U_{xB}$, ($x=x_k$, $k>K$) if the chosen thresholds ($t_x$, $x=x_k$, $k>K$) take the extreme value $t_x \geq 1$ if $U_{xB} < ½$ and $t_x < 0$ if $U_{xB} > ½$. Then, in each state of nature s, the expression $(1-U_{xB})\cdot 1[F(x,b_s)>t_x] + U_{xB}\cdot 1[F(x,b_s)\leq t_x]$ in (24) equals $U_{xB}$ when $U_{xB} < ½$ and equals $1-U_{xB}$ when $U_{xB} > ½$. Hence, MR equals the lower bound in (25). Note that this lower bound is less than MMR with no data, because the summation in (25) is over a subset of x values.

One might think use of extreme thresholds to be unintuitive. The rationale is clear when one recalls that the weighted predictors considered in this section are uninformative about $P(y=1|x)$ for ($x=x_k$, $k>K$). The only available information to guide treatment choice for these values of x is the value of the welfare $U_{xB}$.

We showed in Section 2.2 that, if the state space is unrestricted, MR is minimized by assigning treatment A when $U_{xB} < ½$ and by assigning B when $U_{xB} > ½$. In the borderline case with $U_{xB} = ½$, MR is invariant to treatment choice. Use of extreme thresholds yields these treatment choices.

To conclude, we summarize the main findings in this section. Suppose that one uses plug-in prediction as discussed here. Then it is best from the perspective of maximum regret to select thresholds as follows: $t_x = 1 - U_{xB}$ for $(x = x_1, \ldots, x_K)$; $t_x \geq 1$ if $U_{xB} < ½$ and $t_x < 0$ if $U_{xB} > ½$ for $x_k$, $k>K$. If $U_{xB}$ is x-invariant for $(x_k, k>K)$, it is best to let $x_1, \ldots x_K$ be the K values of x with the highest covariate probabilities.

3.3.1. Applications to Extrapolation from a Single Sub-Population

A simple and realistic class of applications of the above analysis occurs when K=1. Consider the common situation in which empirical research studies prediction only for persons with covariate value $x_1$ and yields correct predictions for this sub-population. Outcomes are not observed for persons with other covariate values. Lacking evidence, a decision maker assumes that the event probability does not vary with the value of x. Thus, the decision maker uses the plug-in predictor $\varphi_{sx(k)} = p_{sx(1)}$, $k > 1$. Then the MR findings that we have reported hold.

## 4. Numerical Computation of Limit and Finite-Sample Maximum Regret for Plug-In Predictors

4.1. Scope of Study

The algebraic analysis of Sections 3.1, 3.2, and 3.3 is instructive but limited in scope. This section reports computational findings that broaden our understanding of the limit and finite-sample MR of plug-in prediction. We use computation to generalize our knowledge of limit MR using the marginal probability and parametric prediction models, as well as to provide a sense of the adequacy of the limit approximation and evidence on the implications of statistical imprecision for both misspecified and correctly specified prediction models.

Section 3.1 characterized limit MR with a correctly specified prediction model. We include as a predictor in this computational study the nonparametric (cell probability) predictor $\varphi_{sx} = p_{sx}$ in the limit case and, analogously, the cell proportion $\frac{1}{N_x}\sum_i y_i$ · $1[x_i = x]$ in the finite-sample case, where $N_x = \sum_i 1[x_i=x]$. We use these predictors as benchmarks against which to assess the computational results with misspecified predictors, as well as to explore limit MR of a correctly specified predictor with suboptimal thresholds.

Section 3.2 characterized the performance of the marginal probability predictor only when x is binary and a particular predictor threshold is used to choose treatments. Assuming knowledge of the marginal probability $P(y=1)$ and the covariate distribution $P(x)$, Cross and Manski (2002) studied joint identification of $P(y=1|x)$, $x \in X$ when X is any finite set. They showed that the joint identification region is a bounded convex set whose extreme points are the expectations of a vector of certain *stacked distributions*. These distributions are complex when the cardinality of X is greater than two, making algebraic analysis of limit MR infeasible. However, computational study is tractable.

Section 3.3 characterized limit MR with an invertible K-dimensional parametric model, choosing a parameter value to correctly predict K conditional probabilities. Applied researchers rarely use such an estimation method when data are available on (y,x) for k>K values of x. The prevalent practice is instead to choose a parameter value that minimizes some measure of distance from predictions to observed outcomes. It is particularly common to minimize the squared deviation between predictions and outcomes. When the data $\psi$ are a random sample of size N drawn from P(y,x), one solves the problem

$$(28) \quad \min_{b \in B} \; 1/N \sum_{i=1,\ldots,N} [y_i - F(x_i,b)]^2.$$

As $N \rightarrow \infty$, the limiting form of this problem in state of nature s is

$$(29) \quad \min_{b \in B} \sum_{x \in X} P(x)[p_{sx} - F(x,b)]^2.$$

The summation in (29) is the mean square error (MSE) of prediction with parameter value b. Let $b_s$ denote a solution to this problem, which minimizes MSE. Using our notation for limiting predictors, $\varphi_{sx} = F(x,b_s)$.

Comparing MR and MSE, both criteria weight losses by the covariate distribution in the same way. An important difference, however, is how the deviation between $p_{sx}$ and $\varphi_{sx}$ degrades the criterion. With MSE, what matters is the distance between $p_{sx}$ and $\varphi_{sx}$, symmetrically for positive and negative deviations. For MR, what matters is whether the two are on the same side of their thresholds, $1-U_{xB}$ and $t_x$, respectively. If $p_{sx}$ is less (greater) than $1-U_{xB}$, negative (positive) deviations have no harmful consequences in terms of MR. Positive (negative) deviations are harmful if they are sufficiently large to cross the threshold. The extent of degradation increases with $|p_{sx}-(1-U_{xB})|$. Note that $U_{xB}$ plays no role in MSE.

Algebraic analysis of the limit MR of plug-in prediction using limit predictors that minimize MSE or the misclassification rate appears intractable, but we can approximate limit and finite-sample MR numerically. We present findings for the familiar cases in which the prediction model is linear in parameters or a logit model or a binary linear classifier. These are special cases of (29) where F(x,b) has the following forms: $F(x,b)=xb$ or $F(x,b)=\exp(xb)/[1+\exp(xb)]$ or $F(x,b)=1[xb>0]$, respectively.

## 4.2. Experimental Design

We consider a setting where the set of covariate values is $X = \{0,1,2,3\}$ with a uniform covariate distribution, placing probability ¼ on each value. We initially assume that $U_{xB}$ is x-invariant, taking the value 0.2 or 0.5 for all values of x. We then consider x-specific $U_{xB}$.

In the absence of any a priori knowledge of P(y|x), we ideally would want the state space S to encompass all elements of the four-dimensional unit hypercube $[0,1]^4$. In computational practice, we approximate the cube by a four-dimensional grid, in which each conditional probability can take one of finitely many values. In our design, the x-

specific grid for $p_{sx}$ permits the twenty-one values [0,0.05,0.1,0.15,…,1]. Hence, the approximate state space contains $21^4$ = 194,481 elements.

In principle, we would want the feasible probabilistic prediction thresholds for treatment choice to encompass all elements of the cube $[0,1]^4$, enabling the planner to choose any vector $(t_0,t_1,t_2,t_3)$ of covariate-specific thresholds. In practice, we consider nine thresholds $t_x$ = 0.1,0.2,0.3,…,0.9 for each value of x, yielding a total of $9^4$ = 6,561 different threshold vectors. We do not consider extreme thresholds 0 and 1, which would make plug-in prediction superfluous.

Given a specified threshold vector, we compute maximum regret by grid search over the 194,481 grid elements of S. With 6,561 thresholds, this requires 194,481 x 6,561 = 1,276,989,841 regret computations for each predictor and specification, as well as for each finite-sample simulation. Each regret computation is simple for the marginal probability and linear predictors. Logit regret computation requires non-negligible computation time because it requires solution of a nonlinear least squares problem in each state. Computations were performed in Python, using the trust region reflective optimization algorithm in *scipy.optimize.least_squares* from the SciPy library for nonlinear least squares.

Minimizing classification error also requires non-negligible computation time because it requires finding the minimum of a step function. The resulting binary classifier is often viewed as non-probabilistic, but it can be viewed as a probabilistic prediction that only takes the values 0 or 1.

This extremum problem has been studied in the context of maximum score estimation, which is algebraically equivalent to minimizing classification error. In general, finding the maximum is complex. However, Manski and Thompson (1986) showed how the problem simplifies for our specification with scalar x or, more generally, a monotone linear index. The median regression version of maximum score prediction was implemented by an exact finite maximization of the population score over the normalized parameter space, using the great-circle approach presented by Manski and Thompson. By construction, treatment assignment with the binary classifier is invariant to the choice of threshold.

The above procedures may be used to approximate both limit and finite-sample MR. The only difference between these cases is ease of computation of the state-dependent error probability of $Q_s\{e[p_{sx}, \varphi_x(\psi), t_x, U_{xB}] = 1\}$ of equation (16). In the limit, the error probability takes only the values 0 and 1 and is simple in form, as shown in Section 3. In finite samples, the error probability can take any value in the [0,1] interval and, in general, must be approximated by Monte Carlo simulation.

To approximate the finite-sample error probabilities, we perform 50 Monte Carlo simulations for each state $p_{sx}$, where each simulation draws 10 realizations of $y_i$ from a Binomial $(10,p_{sx})$ distribution for each of the four values of x, yielding a total sample of size N=40 in each case. Each draw of 40 observations is used to calculate each prediction of $p_{sx}$ and then determine the treatment assignment error at each value of x for each prediction-threshold pair. For each value of x and prediction threshold-pair, the error is either 0 or 1 in any one simulated draw of a sample from $B(10,p_{sx})$. These errors are averaged across the 50 simulated draws to approximate the state-dependent error probabilities of equation (16). Note that the cell proportion predictor offers an instructive exception to the need for approximation by Monte Carlo simulation, as the error probability in a given state s can be calculated exactly based on the known Binomial distribution of the sample data for each value of x.

As described above, the study begins without any restrictions on the state space, approximated by a grid of $21^4$ values on the four-dimensional unit hypercube. We then repeat the computations restricting the state space to weakly increasing functions of x, such that $p_{s0} \leq p_{s1} \leq p_{s2} \leq p_{s3}$.

### 4.3. Findings

In this section, we summarize the computational results and highlight key findings. Supplemental Appendix C contains detailed results and findings. We begin with the unrestricted state space, and then present findings with the monotonicity restriction on the state space. In each case, we first discuss limit MR and then assess finite-sample MR to provide a sense of the adequacy of the limit approximation and evidence on the implications of statistical imprecision for threshold choice and treatment choice.

4.3.1. Specifications with the Unrestricted State Space

*Limit MR*

Table 2 summarizes the computational results for limit MR in two of the four specifications of welfare, one with x-invariant welfare and one with x-specific welfare. The initial specification (top three rows) sets x-invariant welfare equal to 0.5. We showed in Section 2.2 that, in the absence of data, MR is invariant to treatment choice with $U_{xB}$=0.5; in fact, MR=0.5 for each value of x. Thus, we report in the second column of Table 2 that singleton MMR with no data equals 0.50.

Table 2. Limit Maximum Regret with the Unrestricted State Space

| Welfare | Singleton Minimax Regret with No Data | Threshold | **Limit Maximum Regret** | | | | |
|---|---|---|---|---|---|---|---|
| | | | Marginal Probability | Linear Prediction Square Loss | Logit Prediction Square Loss | Min MCR | Cell Probability |
| $U_{xB}$ = 0.5 | | x-Invariant t | 0.25 | 0.30 | 0.26 | | 0.00 |
| for all x | 0.50 | $t_x = 1 - U_{xB}$ | 0.25 | 0.34 | 0.31 | 0.13 | 0.00 |
| | | x-Specific $t_x$ | 0.25 | 0.29 | 0.26 | | 0.00 |
| $U_{xB}$ = | | x-Invariant t | 0.40 | 0.44 | 0.43 | | 0.28 |
| [0.1, 0.3, 0.7, 0.9] | 0.20 | $t_x = 1 - U_{xB}$ | 0.20 | 0.20 | 0.23 | 0.31 | 0.00 |
| | | x-Specific $t_x$ | 0.20 | 0.20 | 0.23 | | 0.00 |

The first row of entries for this specification, labelled *x-invariant t*, reports limit MMR for each prediction method constraining the thresholds to be x-invariant, taking on any one of the nine values t = 0.1, 0.2, 0.3, …, 0.9. Among all misspecified prediction models, the minimum MCR predictor minimizes MR, yielding the value 0.13 for any threshold, as reported under *Min MCR*.[4] We see here that constrained minimum MMR is at least 0.25 for each of the other misspecified predictors. Section 3.1.1 showed that ½ is the optimal threshold with a correctly specified predictor in this specification, yielding limit MR equal to zero. This is reported in the final column, labelled "cell probability," for the nonparametric predictor that takes limit value $p_{sx}$ for each x.

The optimal threshold for a correctly specified predictor—that is, $t_x = 1 - U_{xB}$—need not perform well when a misspecified plug-in predictor is used. The second row of entries, labelled $t_x = 1 - U_{xB}$, makes this point clear, as it reports MR for each predictor using the threshold value t = ½. Limit MR rises relative to the first row of entries with x-invariant t for both the linear predictor and the logit predictor, where the minimum arises at $t \in \{0.3, 0.7\}$ and t = 0.6, respectively. Recall that treatment assignment, and hence MR, with the binary classifier is invariant to the choice of threshold.

Contrariwise, a correctly specified predictor need not perform well when a suboptimal threshold is used. As detailed in Supplemental Appendix C, while MR for the cell probability predictor is zero at the optimal t=½, it increases with the distance from the optimal threshold up to 0.40 at t=0.9, close to MMR with no data.

The third row of entries, labelled *x-specific t*, reports MMR for each prediction method constraining the vector of thresholds $t_x$ to take on any of the $9^4$ = 6,561 combinations of the nine values from 0.1 to 0.9. With x-invariant welfare, limit MMR exhibits a small decline (0.01) when allowing for x-specific thresholds with the linear predictor, where $t_x$ = [0.3,0.4,0.6,0.7], relative to MR with x-invariant t. MMR does not decline for any other predictor in this specification of welfare.[5]

The bottom three rows of Table 2 pertain to the specification with x-specific welfare where $U_{xB}$ = [0.1,0.3,0.7,0.9]. Here, singleton MMR with no data equals 0.20, which is considerably lower than limit MMR for any prediction method paired with an x-invariant threshold, ranging from 0.28 for the cell probability to 0.40 or more for the marginal probability and the linear and logit predictors. With one exception, relaxing the constraint that thresholds be x-invariant—in particular, setting $t_x = 1 - U_{xB}$—reduces limit MMR by about half for the misspecified predictors, while limit MMR falls to zero for the correctly specified predictor. The aforementioned exception is the $t_x$-invariant binary classifier, with limit MR of 0.31 in this specification. These findings demonstrate the potentially harmful effects on welfare of using misspecified plug-in predictors and decision thresholds for treatment choice.

[4] In contrast, in the specification with $U_{xB}$=0.2, MR=0.30 for the minimum MCR, double the MMR of the marginal probability and logit predictors with x-invariant t.

[5] In the specification with $U_{xB}$=0.2, MMR does not decline for any of the predictors.

*Finite-Sample MR*

Table 3 summarizes the finite-sample results for these two specifications. Two striking findings arise from the full set of results presented here and in Supplemental Appendix C. First, the results are typically close to those for limit MR. This suggests that our analysis of limit MR in Section 3 provides a useful approximation to finite-sample MR when sample size is only moderate. Second, a notable deviation between the findings for limit and finite-sample MR is that the latter is sometimes smaller than the former when the prediction-threshold pair is suboptimal.

Table 3. Finite-Sample Maximum Regret with the Unrestricted State Space

| Welfare | Singleton Minimax Regret with No Data | Threshold | **Finite-Sample Maximum Regret** | | | | |
|---|---|---|---|---|---|---|---|
| | | | *N = 40*, 50 simulations per specification | | | | |
| | | | Marginal Proportion | Linear Prediction Square Loss | Logit Prediction Square Loss | Min MCR | Cell Proportion |
| $U_{xB} = 0.5$ | | x-Invariant t | 0.25 | 0.25 | 0.25 | | 0.04 |
| for all x | 0.50 | $t_x = 1 - U_{xB}$ | 0.25 | 0.25 | 0.25 | 0.14 | 0.04 |
| | | x-Specific $t_x$ | 0.25 | 0.25 | 0.25 | | 0.04 |
| $U_{xB} =$ | | x-Invariant t | 0.40 | 0.34 | 0.32 | | 0.18 |
| [0.1, 0.3, 0.7, 0.9] | 0.20 | $t_x = 1 - U_{xB}$ | 0.20 | 0.20 | 0.20 | 0.24 | 0.04 |
| | | x-Specific $t_x$ | 0.20 | 0.19 | 0.17 | | 0.04 |

Results where limit MR exceeds finite-sample MR may initially appear surprising, but they are not anomalous. This pattern may perhaps be related to the fact that, in some settings with partial identification, controlled randomization of choice is known to be a feature of minimax regret decision making (Manski, 2009, 2025; Montiel Olea, Qiu, and Stoye, 2026). Manski (2025) observed that basing decisions on random samples of finite size yields a form of uncontrolled randomization. We conjecture that uncontrolled randomization may sometimes approximate the benefit of controlled randomization.

When using a correct prediction model and the optimal threshold, limit MR is zero, hence necessarily less than the positive MR stemming from statistical imprecision of prediction in finite samples. For example, limit MR for the correctly specified cell probability predictor is 0 and finite-sample MR is 0.04 when $U_{xB}=0.5$ and $t_x=0.5$. However, this inequality need not hold when using a misspecified prediction model or a suboptimal threshold such that limit MR is positive. Then sampling imprecision can reduce MR by reducing the frequency of the errors that occur when limit predictions are on the wrong side of the threshold. [6]

4.3.2. Specifications with State Space Restricted to Increasing Functions of x

Here we suppose that the conditional event probability is known to be a weakly increasing function of x.

*Limit MR*

Table 4 shows that the state-space restriction generates notable decreases in limit MR, not only for the predictors that are monotone functions of x but also for the marginal probability predictor with thresholds that vary with x. Further, the binary classifier assigns treatment optimally when $U_{xB}=0.5$.

Linear and logit prediction behave similarly in each specification, yielding minimum MR between 0.04 and 0.08 with $t_x=1-U_{xB}$ in the two specifications with results reported in Table 4, as well as between 0.04 and 0.10 in the two specifications not reported here. Limit MMR is also less than 0.10 for the marginal probability with x-specific thresholds. For example, with x-invariant $U_{xB}$=0.5, MMR=0.08 for this predictor paired with $t_x$ = [0.8,0.6,0.4,0.2], as opposed to MMR=0.25 with $t_x = 1-U_{xB} = 0.5$. In the two specifications with x-specific welfare, MMR obtains for the marginal probability with $t_x$ = [0.9,0.7,0.3,0.1].

[6] For example, consider the cell probability predictor when $U_{xB} = 0.5$, $t = 0.1$, and $p_{sx} = 0.15$. The optimal threshold is 0.5 and the optimal treatment is A because $0.15 < 0.5$. The limit error probability using threshold 0.1 is 1 because $0.15 > 0.1$, implying that treatment B is wrongly assigned in the limit. Yet the finite sample error probability may be much less than one. In our computations, sample size with each value of x is 10. When sample data are generated by the Binomial (10, 0.15) distribution, the finite-sample error probability is just 0.46.

Table 4. Limit Maximum Regret with the Restricted State Space

| Welfare | Singleton Minimax Regret with No Data | Threshold | **Limit Maximum Regret** | | | | |
|---|---|---|---|---|---|---|---|
| | | | Marginal Probability | Linear Prediction Square Loss | Logit Prediction Square Loss | Min MCR | Cell Probability |
| $U_{xB}$ = 0.5 | | x-Invariant t | 0.25 | 0.08 | 0.05 | | 0.00 |
| for all x | 0.50 | $t_x = 1 - U_{xB}$ | 0.25 | 0.08 | 0.05 | 0.00 | 0.00 |
| | | x-Specific $t_x$ | 0.08 | 0.08 | 0.05 | | 0.00 |
| $U_{xB}$ = | | x-Invariant t | 0.40 | 0.15 | 0.14 | | 0.15 |
| [0.1, 0.3, 0.7, 0.9] | 0.20 | $t_x = 1 - U_{xB}$ | 0.08 | 0.08 | 0.04 | 0.15 | 0.00 |
| | | x-Specific $t_x$ | 0.08 | 0.08 | 0.04 | | 0.00 |

*Finite-Sample MR*

Table 5. Finite-Sample Regret with the Restricted State Space

| Welfare | Singleton Minimax Regret with No Data | Threshold | **Finite-Sample Maximum Regret** | | | | |
|---|---|---|---|---|---|---|---|
| | | | *N = 40* , 50 simulations per specification | | | | |
| | | | Marginal Proportion | Linear Prediction Square Loss | Logit Prediction Square Loss | Min MCR | Cell Proportion |
| $U_{xB}$ = 0.5 | | x-Invariant t | 0.25 | 0.06 | 0.04 | | 0.04 |
| for all x | 0.50 | $t_x = 1 - U_{xB}$ | 0.25 | 0.06 | 0.04 | 0.03 | 0.04 |
| | | x-Specific $t_x$ | 0.06 | 0.06 | 0.04 | | 0.03 |
| $U_{xB}$ = | | x-Invariant t | 0.40 | 0.16 | 0.15 | | 0.16 |
| [0.1, 0.3, 0.7, 0.9] | 0.20 | $t_x = 1 - U_{xB}$ | 0.05 | 0.08 | 0.03 | 0.15 | 0.02 |
| | | x-Specific $t_x$ | 0.05 | 0.06 | 0.03 | | 0.02 |

Finite-sample MMR findings are summarized in Table 5. The patterns of variation are very similar to the patterns for limit MR, both qualitatively and quantitatively. As we found with the unrestricted state space, finite-sample MR is often smaller than limit MR when misspecified predictors or suboptimal thresholds are used.

## 5. MR Analysis with Plug-In Estimation of $U_{xB}$

In Sections 2 through 4 the empirical problem motivating MR analysis was incomplete knowledge of the x-specific event probabilities $p_x$. The planner was assumed to know x-specific welfare $U_{xB}$. A further empirical problem may be incomplete knowledge of welfare. Assumptions and sample data may be used to point-estimate these unknowns as well, thereby extending the scope for plug-in decision making. For example, health economists commonly use stated-preference data obtained in discrete-choice experiments to estimate health-related welfare on a [0,1] quality-of-life scale; see Devlin et al. (2022) and EuroQol Research Foundation (2023). In ML research, Zadrozny and Elkan (2001) proposed a method for cost-sensitive learning where both costs (or welfare) and probabilities are unknown.

The mathematical structure of the treatment-choice problem studied in this paper is simple enough that our analysis can be applied with small modification to settings with incomplete knowledge of welfare. Recall that, given a prediction-threshold pair, a treatment error occurs in state s with sample data $\psi$ if $[p_{sx} \le 1 - U_{xB}, \varphi_x(\psi) > t_x]$ or if $[1 - U_{xB} < p_{sx}, \varphi_x(\psi) \le t_x]$. These inequalities may equivalently be written $[p_{sx} + U_{xB} \le 1, \varphi_x(\psi) > t_x]$ or if $[1 < p_{sx} + U_{xB}, \varphi_x(\psi) \le t_x]$. Thus, what we regarded earlier as an optimal x-specific threshold $1 - U_{xB}$ for the event probability may equivalently be viewed as an optimal x-invariant threshold 1 for the sum of the x-specific event probability and welfare.

When welfare is not known, a planner may choose treatments by combining plug-in prediction of $p_x$ with plug-in estimation of $U_{xB}$. Extend the notation used earlier so that S encompasses the feasible values of $(p_x, U_{xB}; x \in X)$. Let $\psi$ include the sample data used to estimate welfare. Let $\upsilon_{xB}(\psi)$ be an estimate of $U_{xB}$ and Y be a set of feasible welfare estimators. Let $\tau_x \in [0,2]$ be a threshold applied to the plug-in sum $\varphi_x(\psi)+\upsilon_{xB}(\psi)$. Then a treatment error occurs in state s if $[p_{sx}+U_{xB}\leq 1,\ \varphi_x(\psi)+\upsilon_{xB}(\psi) > \tau_x]$ or if $[1<p_{sx}+U_{xB},\ \varphi_x(\psi)+\upsilon_{xB}(\psi)\leq\tau_x]$.

Extending the earlier equation (16), a predictor function, welfare estimator, and thresholds minimize maximum regret if they solve the problem

(30)
$$\min_{\varphi\in\Phi,\ \upsilon\in\Upsilon,\ \tau_x\in[0,2],\ x\in X}\ \max_{s\in S} \sum_{x\in X} P(x)\,|(1-U_{xB})-p_{sx}|\cdot Q_s\{e[p_{sx},\ \varphi_x(\psi),\upsilon_{xB}(\psi),\tau_x,U_{xB}]=1\}.$$

The limit regret problem (17) extends similarly.

## 6. Conclusion

This paper sheds new light on limit MR and sample MR when plug-in prediction is performed with misspecified prediction models or suboptimal decision thresholds are used. We report informative algebraic analysis of limit MR when a planner applies either correct prediction models with suboptimal thresholds or elementary versions of canonical misspecified prediction models. Our computational study demonstrates, both in the limit and in finite samples, how the quality of decision making can suffer when the optimal prediction-threshold pair is not used.

We see two clear paths for future work. An immediate direction would be to assess more complex prediction models, such as those with high-dimensional predictors. The major obstacle here is computation time. A more intriguing direction would address the prediction-threshold problem by jointly choosing a prediction model and thresholds within a specified class of models and thresholds, as an alternative to relying on plug-in prediction methods. We are currently engaged in this research.

## Supplemental Appendix A: Related Literature on Prediction-Threshold Problems

We observed in the text that the treatment problem studied in this paper arises in many contexts and that plug-in prediction is a common decision strategy. There is a large and diverse related literature, from which our analysis differs in two primary respects. Whereas we study joint choice of probabilistic predictions and decision thresholds, it has been common to view these choices sequentially. A prevalent practice has been to first choose a prediction method using some criterion that measures quality of prediction, without reference to a specific decision problem where the prediction will be used. After a prediction method has been chosen, decision thresholds are selected. Sadana et al. (2025) recently reviewed exceptions to this characterization of plug-in prediction research in ML and operations research, labelled *data-driven optimization*, which seeks to address both prediction and decision making via "sequential learning and optimization" (SLO) or "integrated learning and optimization" (ILO).

While the literature proposes many criteria to evaluate predictions, we are not aware of MR analysis as we perform here. When regret is discussed, it is typically ex post experienced regret rather than ex ante regret. Moreover, the focus has been on the zero regret achievable with correct predictors. See, for example, Hart and Mas-Collell (2000) and Foster and Hart (2018). There are some recent exceptions that are concerned with versions of ex ante regret, but the definition of regret differs notably from ours in each case, as well as from each other. For example, Elmachtoub et al. (2025) compared the regret of SLO and ILO optimization methods but, analogous to Kitagawa and Tetenov (2018), they computed regret relative to a constrained optimum within a potentially misspecified parametric family of distributions. Rothblum and Yona (2023) used a concept of ex ante regret that permits regret to be negative.

This appendix summarizes and critiques the literature. We begin with work that views prediction as a self-contained objective for research. We then turn to research that uses predictions to make decisions.

A.1. Measurement of Prediction Performance

Frequentist statisticians typically measure prediction performance by mean square error across repeated samples, without reference to a subsequent decision problem. The machine learning practice of minimizing the overall misclassification rate with binary outcomes and predictions is equivalent to minimizing mean squared error

A subset of machine learning research considers minimization of the overall misclassification rate to be inappropriate if the marginal distribution of outcomes is deemed "imbalanced," in the sense that P(y=1) is distant from ½. We are unaware of research that shows concern for balance conditional on observed covariates.

When imbalance is found, it has often been recommended to artificially balance the available data by underweighting the majority outcome and/or overweighting the minority outcome. The influential SMOTE proposal of Chawla et al. (2002) does both. To motivate their weighting proposal, the authors wrote (p. 322):

> "The performance of machine learning algorithms is typically evaluated using predictive accuracy. However, this is not appropriate when the data is imbalanced and/or the costs of different errors vary markedly."

We are perplexed that neither these authors nor others advocating weighting data provide analysis explaining why they view imbalance to be a problem and how SMOTE or other types of weighting can be beneficial.

In contemporaneous work, Elkan (2001) derived conditions under which weighting would accomplish the same objective as cost-sensitive learning. However, Elkan concluded (p. 973): "The recommended way of applying one of these methods in a domain with differing misclassification costs is to learn a classifier from the training set as given, and then to compute optimal decisions explicitly using the probability estimates given by the classifier."

Yet, the SMOTE methodology and other weighting methods have become widely cited and utilized. We have found one recent article questioning the premise that imbalance poses a problem. Van den Goorbergh et al. (2022) conclude (p. 1522):

"Outcome imbalance is not a problem in itself, imbalance correction may even worsen model performance."

A separate strand of research in engineering and machine learning advocates use of the receiver operating curve (ROC) to measure prediction performance. These investigators sometimes begin with a decision problem reminiscent of Pauker and Kassirer (1975), but they state that their objective is to measure prediction performance in the absence of knowledge of the welfare losses due to different treatment errors and, relatedly, without specifying a decision threshold. See, for example, Bradley (1997), Provost and Fawcett (2001), Chawla et al. (2002), and Bach, Heckerman, and Horvitz (2006).

The ROC is a graphical representation of the prediction performance of a binary classifier derived from a probabilistic prediction model for a binary event y conditional on x. Let $\varphi(x)$ be a probabilistic prediction of y. Its value may be the same or differ from the true probability $P(y=1|x)$. Let t be a specified threshold and let the binary classification of y based on $\varphi(x)$ and t be 1 if $\varphi(x) > t$ and 0 if $\varphi(x) \leq t$. The *true positive rate* (*sensitivity*) of the prediction is defined to be $P[\varphi(x)>t|y=1]$ and the *false positive rate* (1 – *specificity*) is $P[\varphi(x >t|y=0]$. The ROC computes these values for all $t \in [0,1]$ and plots the true positive rate against the false positive rate. Both values weakly decrease in t, so the true positive rate weakly increases with the false positive rate.

Our reading of research studying the ROC indicates that authors use the curve in two ways. One is to provide a simple graphical expression of the tradeoff between the true and false positive rates that occurs as the threshold varies. Each point on the curve is attainable for some value of t, so selecting a point on the curve is implicitly equivalent to selecting a threshold. The other use of the ROC is to provide a scalar summary of prediction performance, one that is not threshold dependent. It has been common to compute the *area under the curve* (AUC).

We are unaware of any use of the ROC that coherently informs the prediction-threshold problem studied in this paper. The true and false positive rates, as defined in the ROC, are conditional probabilities that are not relevant to treatment choice. The relevant probability is P(y=1|x). If one wants to define true and false positive rates based on P(y = 1|x), they would be P[y=1|x, φ(x)>t] and P[y=0|x, φ(x)>t] respectively. The irrelevance of the ROC to clinical decision making has been recognized in medical research. See Altman and Bland (1994), Feinstein (2002), and Manski (2021). Nevertheless, study of the ROC continues to be prominent in medicine and elsewhere.

A.2. Selection of Decision Thresholds

Researchers have occasionally recognized verbally that predictions used to inform decision making should be evaluated by reference to the relevant decision problem. The conclusion to the Kleinberg et al. (2018) study of predicting criminality and judicial bail decisions called attention to the common sequential process of choice of a prediction method followed by consideration of decision making, writing (p. 288):

> "The bail example highlights the value of solving social science problems of the type: Data → Prediction → Decision. Machine-learning applications typically focus solely on the Data → Prediction link. The objective is to search through different candidate prediction functions and identify the one with the greatest prediction accuracy—a "bake-off." Algorithm performance tends to be quantified on predictive value. The bail example, though, illustrates why understanding the Prediction → Decision link is at least as important. . . . . Good predictors do not necessarily improve decisions."

These remarks recognize the problem with the standard practice, but the authors did not propose a solution. In particular, they do not mention statistical decision theory as a potential framework for evaluation of prediction methods. In a recent paper, Babii et al. (2006) moves partly in this direction by studying the asymptotic performance of empirical risk minimization.

Researchers studying choice of decision thresholds have advocated an array of approaches that vary in their declared motivations and justifications. Although Pauker and Kassirer (1975) derived the optimal threshold when event probabilities and patient utilities are known, our search of the subsequent literature indicates that threshold choice in practice has generally not been grounded in their finding.

It has been common to recommend choice of treatment B if the estimate of P(y=1|x) > ½ and A otherwise. Esposito et al. (2021) wrote (p. 2623): "For binary classification, instances with a positive probability ≥ 0.5 are typically predicted as positives, otherwise as negatives." This practice should be regarded as a convention rather than based on a clear principle. Using the threshold ½ is well motivated from the perspective of MR if the losses in welfare stemming from the two types of treatment error (choosing B when A is better and vice versa) are identical and if the prediction model is correctly specified. However, justification is generally absent outside of these special circumstances. Arguing that imbalance provides a reason to use a threshold other than ½, Esposito et al. discussed numerous algorithmic criteria for threshold selection that they asserted should mitigate imbalance. Given that we do not understand why machine learning researchers consider imbalance to be problematic, we are unable to motivate these criteria.

In medical settings, there is often a heuristic appreciation that the losses in patient welfare from the two types of treatment error are not identical. When this is so, panels setting clinical practice guidelines (CPGs) generally recommend thresholds different from ½. Prominent examples are CPGs for prophylactic treatment of currently healthy women assessed to be at risk of developing breast cancer. As noted in the Introduction, it has been standard in the United States to use the Breast Cancer Risk Analysis (BCRA) Tool of the National Cancer Institute to predict the probability that a woman who currently does not suffer from breast cancer will develop the disease in the next five years. National Comprehensive Cancer Network (2026) recommends annual surveillance if the predicted probability is below 0.017 and some form of prophylactic treatment if the probability is higher. We do not find in the documentation an explanation of the rationale for use of the x-invariant threshold 0.017. We think it curious that the threshold does not vary with such patient attributes as age and comorbidities, which may affect the welfare losses from treatment errors. Patel and Dominitz (2024) describe age-group variation in colorectal cancer screening recommendations that imply higher decision thresholds for ages 76 to 85 than for ages 50 to 75. Further, recognizing the "competing mortality risks, the generally long lag time in the progression from precancerous polyps to cancer, and the greater risk for complications in elderly persons, any screening benefits are likely to be outweighed by harms for persons older than 85 years." That is, $t_{age > 85} = 1.0$.

Supplemental Appendix B: Maximum Regret for Binary x with Prediction by the Marginal Mean and x-Invariant Threshold

Let x be binary, with values x=0 and x=1. Let $p_s \equiv P(x=0)p_{s0} + P(x=1)p_{s1}$ be the predictor of $p_{sx}$. Let $t^* \equiv P(x=0)(1-U_{0B}) + P(x=1)(1-U_{1B})$ be the x-invariant threshold

First we show that, with predictor $p_s$ and threshold $t^*$, treatment is optimally assigned for at least one value of x. To do so, we must consider two cases: (I) $p_s \leq t^*$, so treatment A is assigned for both values of x and (II) $p_s > t^*$, so treatment B is assigned for both values of x.

I. Suppose that $p_s \leq t^*$, implying that A is assigned for x = 0 and x = 1.
   a. Suppose that $p_{s1} > (1-U_{1B})$, implying that B is optimal for x = 1. Rewrite $p_s \leq t^*$ as $P(x=0)p_{s0} + P(x=1)p_{s1} \leq P(x=0)(1-U_{0B}) + P(x=1)(1-U_{1B})$. For $p_s \leq t^*$ to hold with $P(x=1)p_{s1} > P(x=1)(1-U_{1B})$, it must be that $p_{s0} < (1-U_{0B})$, Hence, A is optimal for x=0.
   b. Suppose that $p_{s0} > (1-U_{0B})$, implying that B is optimal for x=0. The same logic shows that $p_{s1} < (1-U_{1B})$. Hence, A is optimal for x=1.

II. Suppose that $p_s > t^*$, implying that B is assigned for x=0 and x=1.
   a. Suppose that $p_{s1} \leq (1-U_{1B})$, implying that A is optimal for x=1. Rewrite $p_s > t^*$ as $P(x=0)p_{s0} + P(x=1)p_{s1} > P(x=0)(1-U_{0B}) + P(x=1)(1-U_{1B})$. For $p_s > t^*$ to hold with $P(x=1)p_{s1} \leq P(x=1)(1-U_{1B})$, it must be that $p_{s0} > (1-U_{0B})$. Hence, B is optimal for x=0.
   b. Suppose that $p_{s0} \leq (1-U_{0B})$, implying that A is optimal for x=0. The same logic shows that $p_{s1} > (1-U_{1B})$. Hence, B is optimal for x=1.

Thus, there are four cases with positive regret:

Ia. Assign A to all, but B is optimal for x=1

Ib. Assign A to all, but B is optimal for x=0

IIa. Assign B to all, but A is optimal for x=0

IIb. Assign B to all, but A is optimal for x=1

We determine maximum regret in each case. The finding depends on whether $P(x) > t^*$ or $P(x) \leq t^*$, x = 0, 1. [7] Hence, we need to consider eight settings in all.

We show that in each setting, the maximum (supremum) of regret obtains when the marginal event probability $p_s$ takes (approaches) the threshold value $t^*$, and the conditional means $p_{s0}$ and $p_{s1}$ take opposing extreme values that yield $p_s = t^*$. To determine these extremes, we use the fact that the Law of Total Probability implies the identification regions for $p_{s0}$ and $p_{s1}$ are as follows:

$$p_{s0} \in [0, 1] \cap [\ (p_s - P(x=1))/P(x=0),\ p_s/P(x=0)\ ]$$

$$p_{s1} \in [0, 1] \cap [\ (p_s - P(x=0))/P(x=1),\ p_s/P(x=1)\ ].$$

This result was informally sketched by Duncan and Davis (1953) in their study of ecological inference. Horowitz and Manski (1995) appear to have published the first formal proof in their study of contaminated sampling. Cross and Manski (2002) considered both intervals jointly in their study of short versus long regression.

The Cross and Manski analysis of joint identification of conditional probabilities observed that $p_{s0}$ attains its lower bound when $p_{s1}$ attains its upper bound, and vice versa. This is important to our present analysis. When $p_s$ and P(x) yield an informative lower bound on one of the conditional probabilities $p_{sx}$—that is, the lower bound is positive—they do not yield an informative upper bound on the other conditional probability—that is, the upper bound equals one. Symmetrically, when $p_s$ and P(x) yield an informative upper bound on one conditional probability, the lower bound of the other equals zero. We use these relationships in the proof below.

---

[7] The value of $t^*$ places no restrictions of the sign of this inequality. To see this, consider $P(x = 1) > t^*$, which may be rewritten as $P(x = 1) > [(1 - P(x = 0))(1 - U_{0B}) + P(x = 1)(1 - U_{1B})]$. Algebraic manipulation yields $P(x = 1) > [(1 - U_{0B})/(1 - U_{0B} + U_{1B})]$. With $U_{xB} \in (0, 1)$, x = 0, 1, the right side of the inequality can take any value in (0, 1). Thus, both $P(x = 1) > t$ and $P(x = 1) \leq t$ are feasible. The analogous finding applies with x = 0.

Case Ia: Assign A to all, but B is optimal for x=1

1. Suppose $p_s \leq t^*$, implying that A is assigned for both x=0 and x=1.
2. Suppose that $p_{s1} > (1-U_{1B})$, implying that A is erroneously assigned for x=1.
3. We have shown above that, for $p_s \leq t^*$ to hold with $p_{s1} > (1-U_{1B})$, it must be that $p_{s0} \leq (1-U_{0B})$. Hence, A is optimal for x=0.
4. Thus, MR is

$$\max_{p_{s1} \in (1-U_{1B}, 1]} \quad P(x=1)\,[p_{s1} - (1-U_{1B})]$$

$$\text{s.t.} \qquad p_s = P(x=0)p_{s0} + P(x=1)p_{s1}$$

$$p_s \leq t^*$$

$$t^* = P(x=0)(1-U_{0B}) + P(x=1)(1-U_{1B})$$

Regret is maximized where $p_{s1}$ attains its maximum feasible value. To determine this value, we consider two mutually exclusive and exhaustive events: (i) $P(x=1) > t^*$ and (ii) $P(x=1) \leq t^*$.

*(i) Suppose that $P(x=1) > t^*$.*

We have assumed that $p_s \leq t^*$. Hence, $P(x=1) > p_s$. This implies that the upper bound on $p_{s1}$ is less than one and the lower bound on $p_{s1}$ is zero. It follows that, for a given value of $p_s$, the maximum feasible value of $p_{s1}$ is $p_{s1} = p_s/P(x=1)$, which obtains when $p_{s0} = 0$.

The above shows that $p_{s1}$, and hence regret, is maximized when $p_s$ is maximized. It remains to determine the maximum feasible value of $p_s$ subject to the constraints in the regret maximization problem. With $p_{s0} = 0$ and $p_s = t^*$, each constraint is satisfied and the weak inequality constraint on $p_s$ holds with equality. Hence, MR occurs when $p_s = t^*$.

It remains to evaluate MR. Setting $p_{s1} = t^*/P(x=1)$ in problem (4), MR is

$$P(x=1)[t^*/P(x=1) - (1-U_{1B})] \;=\; t^* - P(x=1)(1-U_{1B})$$

$= P(x=0)(1-U_{0B}) + P(x=1)(1-U_{1B}) - P(x=1)(1-U_{1B}) = P(x=0)(1-U_{0B}).$

(ii) *Suppose that $P(x=1) \leq t^*$.*

We do not know whether P(x=1) is greater than or less than $p_s$. However, we know that, for any value $p_s$, the upper bound on $p_{s1}$ is min $\{1, p_s/P(x=1)\}$. Thus, regret in Case Ia may be rewritten as $P(x=1)\ [\min\{1, p_s/(P(x=1)\} - (1-U_{1B})]$. Regret is strictly increasing in $p_s$ until $p_s = P(x=1)$ and invariant thereafter. Hence, maximum regret equals $P(x=1)\ [1-(1-U_{1B})] = P(x=1)U_{1B}$.

Case Ib: Assign A to all, but B is optimal for x = 0

By analogous argument to Case Ia,

- *Suppose $P(x=0) > t^*$:* Maximum regret is $P(x=1)(1–U_{1B})$.
- *Suppose $P(x=0) \leq t^*$:* Maximum regret is $P(x=0)U_{0B}$.

Case IIa: Assign B to all, but A is optimal for x = 1

1. Suppose $p_s > t^*$, implying that B is assigned for both x=0 and x=1.
2. Suppose that $p_{s1} \leq (1-U_{1B})$ such that B is erroneously assigned for x=1.
3. We have shown above that, for $p_s > t^*$ to hold with $p_{s1} \leq 1 - U_{1B}$, it must be that $p_{s0} > 1-U_{0B}$. Hence, B is optimal for x=0.
4. Thus, maximum regret is

$$\max_{p_{s1}\in[0,1-U_{1B}]} \quad P(x=1)[(1-U_{1B}) - p_{s1}]$$

$$\text{s.t.} \quad p_s = P(x=0)p_{s0} + P(x=1)p_{s1}$$

$$p_s > t^*$$

$$t^* = P(x=0)(1-U_{0B}) + P(x=1)(1-U_{1B})$$

Regret is maximized where $p_{s1}$ attains its minimum feasible value. By the Law of Total Probability, for any value of $p_s$, $p_{s1}$ attains its minimum feasible value where $p_{s0}$ attains its maximum feasible value. To determine this value, we consider two mutually exclusive and exhaustive events: (i) $P(x=0) \leq t^*$ and (ii) $P(x=0) > t^*$.

*(i) Suppose $P(x=0) \leq t^*$*

We have assumed that $p_s > t^*$. Hence, $P(x=0) < p_s$. This implies that the upper bound on $p_{s0}$ is one and the lower bound on $p_{s1}$ is greater than zero. It follows that, for a given value of $p_s$, the minimum feasible value of $p_{s1}$ is $p_{s1} = [p_s - P(x=0)]/P(x=1)$, which obtains when $p_{s0} = 1$.

The above shows that $p_{s1}$ is minimized, and hence regret maximized, when $p_s$ is minimized. It remains to determine the minimum feasible value of $p_s$ subject to the constraints in the regret maximization problem. With $p_s > t^*$, the infimum of $p_s$ is $t^*$, in which case $p_{s1} = [t^* - P(x=0)]/P(x=1)$. Hence, the supremum of regret is

$$\begin{aligned} P(x=1)\{(1-U_{1B}) - [t^* - P(x=0)]/P(x=1)]\} &= P(x=1)(1-U_{1B}) - t^* + P(x = 0) \\ &= P(x=1)(1-U_{1B}) - P(x=0)(1-U_{0B}) - P(x=1)(1-U_{1B}) + P(x=0) \\ &= P(x=0) - P(x=0)(1-U_{0B}) = P(x=0)U_{0B}. \end{aligned}$$

*(ii) Suppose $P(x=0) > t^*$*

$P(x=0)$ may be greater or less than $p_s$. We know that, for any value $p_s$, the lower bound on $p_{s1}$ is $\max\{0, [p_s - P(x=0)]/P(x=1)\}$. Thus, regret may be rewritten as $P(x=1)[(1-U_{1B}) - \max\{0, [p_s - P(x=0)]/P(x=1)\}$. Regret is strictly increasing as $p_s$ decreases until $p_s = P(x=0)$ and is invariant thereafter. Hence, the supremum of regret is $P(x=1)[1-U_{1B})]$.

<u>Case IIb. Assign B to all, but A is optimal for x=0</u>

By analogous argument to Case IIa,

- *Suppose $P(x=1) \leq t^*$:* The supremum of regret is $P(x=1)U_{1B}$.
- *Suppose $P(x=1) > t^*$:* The supremum of regret is $P(x=0)(1-U_{0B})$.

Putting these cases together yields maximum (supremum) regret as given in Table 1.

Supplemental Appendix C: Detailed Findings from Computational Analysis

Table SA1 reports detailed computational results with x-invariant thresholds and x-invariant welfare. Detailed results for specifications with x-specific welfare are reported in Table SA2. Table SA3 summarizes the limit and finite-sample computational results for the full set of welfare specifications. Each specification is described in the first column of each table. Corresponding results with the monotonicity restriction on the state space are reported in Tables SA4, SA5, SA6, respectively. We discuss the full set of computational findings below.

C.1. Specifications with x-Invariant Welfare and Thresholds: Unrestricted State Space

Computations of limit MR with x-invariant welfare and thresholds are reported on the left side of Table A1.[8] The Monte Carlo simulation results for finite-sample MR are reported on the right side of this table.

*Limit MR*

The initial specification sets x-invariant welfare equal to 0.5. We showed in Section 2.2 that, in the absence of data, MR is invariant to treatment choice with $U_{xB}=0.5$; in fact, MR=0.5 for each value of x. Thus, we report in the second column of Table 2 that singleton MMR with no data equals 0.5. Each prediction-threshold pair considered in the remaining columns has limit MR less than 0.5. Shaded entries denote the predictor-specific minimax regret over the nine reported values of x-invariant thresholds. We do not shade entries for MR with the minimum misclassification rate (MCR) predictor because it makes binary predictions of 0 or 1. Hence, in this case, MR does not vary with the value of the threshold t.

[8] In Section 3.1, we showed that limit MR for the correctly specified predictor is $\sum_{x\in X} P(x)\,|(1-U_{xB}) - t|$ for any value of t. Yet the computational results in Table A1 deviate from this calculation by 0.05 for any $t < 1-U_{xB}$. This approximation error arises from our computation of regret on a grid of 21 values for the feasible states $p_{sx} \in [0,1]$. The maximum computational approximation error for the cell probability predictor is equal to the distance between grid points, namely 0.05.

Section 3.1.1 showed that ½ is the optimal threshold with a correctly specified predictor in this specification, yielding limit MR equal to zero. This is reported in the final column, labelled "cell probability," for the nonparametric predictor that takes limit value $p_{sx}$ for each x. Contrariwise, a correctly specified predictor need not perform well when a suboptimal threshold is used. We see in Table SA1 that limit MR with the cell probability predictor is zero at the optimal t=½ and increases with the distance from the optimal threshold up to 0.40 at t=0.9, close to MMR with no data.

Among all misspecified prediction models, the minimum MCR predictor (with any threshold) minimizes MR, yielding the value 0.13. MMR is about twice as large for the marginal probability and the logit predictor, and larger still (0.30) for the linear predictor.

The results are qualitatively different in the second specification with x-invariant welfare $U_{xB}$=0.2. Now the optimal (zero MR) x-invariant threshold with a correctly specified predictor is 0.8, and MMR with no data equals 0.2. The MMR among all misspecified prediction models is 0.15 with either (i) the linear predictor and t = 0.9 or (ii) the marginal probability and $t \in \{0.8,0.9\}$. Whereas the minimum MCR predictor outperformed all other misspecified predictors in the first specification with symmetric losses about $U_{xB}$ = 0.5, here in an asymmetric case it yields MR=0.30, which exceeds the MMR for each of the other prediction methods but logit, as well as MMR with no data.

A curiosity of the results in the second specification is that the logit predictor yields the large limit MMR value 0.40, twice the MMR with no data. In contrast, the logit predictor performs relatively well in the first specification. We lack a definite interpretation of this contrasting behavior, but detailed comparison of MR with the linear and logit predictors is revealing. In each state s, both predictors are monotone functions of x of the two-parameter form $F(x,b_s)$, with the slope parameter and intercept chosen to minimize mean square error. Hence, we anticipated that the two predictors would have similar regret properties. In line with anticipation, the top panel of Table 2 finds close to identical MR for most thresholds when $U_{xB}$=0.5 for all x, with the most notable exception arising where t=0.6. The bottom panel similarly finds close to identical MR for thresholds less than 0.6 when $U_{xB}$=0.2 for all x. However, the performance of the linear and logit predictors in the bottom panel diverges sharply at larger thresholds. MR with linear prediction falls from 0.44 to 0.15 as the threshold increases from 0.4 to 0.9. In contrast, MR with logit prediction remains in the interval [0.40,0.44] for all thresholds greater than or equal to 0.4.

In Section C.3 we will find that linear and logit prediction behave similarly in all (welfare, threshold) specifications when the state space is constrained to contain only conditional event probabilities that are weakly increasing in x. Combining this evidence with the findings reported here, we conjecture that the issue with logit performance arises from the conjunction of three potential factors. One is that an unrestricted state space contains states in which the true conditional event probability is far from monotone as a function of x. However, this cannot be the sole explanation as both linear and logit predictors may perform poorly when the true event probability is not monotone in x. The second factor, which differentiates linear and logit prediction, is that the derivative of the former is constant across x whereas, for any parameter vector b, the derivative of the latter approaches zero as F(x,b) goes to zero or one. A third factor, which also differentiates linear and logit prediction, is that the parameter value minimizing mean square error is analytically computable with linear prediction but must be found numerically with logit prediction.[9]

*Finite-Sample MR*

The panel on the right side of Table SA1 reports the finite-sample results. Two striking findings arise. First, the results are typically close to those for limit MR in the left panel. This suggests that our analysis of limit MR in Section 3 provides a useful approximation to finite-sample MR when sample size is only moderate. Second, a notable deviation between the findings for limit and finite-sample MR is that the latter is sometimes smaller than the former.

[9] Concerned that logit performance might be due to nonconvergence of the Python optimization algorithm in some states of nature, we tried a different method—the Levenberg-Marquardt (LM) algorithm rather than the trust region reflective (TRF) algorithm—to minimize mean square error. This yielded somewhat smaller MR values in the suspect parts of Table SA1 but it yielded much larger MR values in the specifications with the restricted state space discussed below, where we anticipated that MR would be relatively small. Further, in unreported results with binary x where limit MR should be zero with the optimal threshold, the TRF algorithm yielded limit MR of zero, whereas the LM algorithm did not. We concluded that the TRF algorithm is more reliable overall.

### C.2 Specifications with x-Specific Welfare and/or x-Specific Thresholds: Unrestricted State Space

Table SA3 summarizes computational results for four specifications of welfare. Two have x-invariant welfare described above ($U_{xB}$=0.2 or $U_{xB}$=0.5 for all x), and two have x-specific welfare ($U_{xB}$ = [0.2, 0.4, 0.6, 0.8] or $U_{xB}$ = [0.1, 0.3, 0.7, 0.9]). Detailed results with x-specific welfare and x-invariant thresholds are reported in Table SA2.

For each specification, Table SA3 reports MMR for each predictor when the thresholds are (i) x-invariant with t taking the nine values reported in Tables SA1 and SA2; (ii) $t_x = 1 - U_{xB}$, which is limit optimal for predictor $\varphi_{sx} = p_{sx}$; or (iii) x-specific with the vector of thresholds taking the $9^4$ = 6,561 combinations of the same nine values. The first row of Table SA3, as well every third row thereafter, is labelled "x-Invariant t MMR" and reports predictor-specific MMR from Table SA1 (Table SA2) for the corresponding specification of x-invariant (x-specific) welfare. The third row, labelled "x-Specific $t_x$ MMR," shows how much MMR is reduced by allowing $t_x$ to vary across x. By construction, treatment assignment with the binary classifier is invariant to the choice of threshold, so allowing $t_x$ to vary across x has no impact on MMR for the minimum MCR predictor.

*Limit MR*

Computations of limit MR are reported on the left side of Table SA3. With x-invariant welfare, limit MMR exhibits a small decline (0.01) when allowing for x-specific thresholds with the linear predictor in the first specification ($U_{xB}$=0.5), relative to MMR constraining thresholds to be x-invariant. MMR does not decline for any other predictor in this specification or in the second specification with x-invariant welfare ($U_{xB}$=0.2). Limit MR is greater with $t_x = 1 - U_{xB}$ than with the MR-minimizing x-invariant threshold for the linear and logit predictors when $U_{xB}$=0.5 and for the linear predictor when $U_{xB}$=0.2.

The pattern of MMR is very different for the specifications with x-specific welfare. Here we obtain striking findings regarding the suboptimality of constraining thresholds to be x-invariant. Consider the fourth specification, with welfare vector $U_{xB}$ = [0.1, 0.3, 0.7, 0.9]. In this case, MMR with no data is 0.20, while MMR with correct cell probability predictions but using an x-invariant threshold is forty percent larger at 0.28. As detailed in Table SA2, this value for limit MR arises with $t \in \{0.3, 0.4, 0.5, 0.6\}$. MMR with misspecified predictors is even higher at each x-invariant threshold value.

The bottom half of Table SA3 shows that, with x-specific welfare, MMR falls when thresholds vary with x for every predictor that is threshold dependent, notably more so for the fourth specification with the greater degree of variation in welfare, relative to the third specification. Even here, however, MMR with no data (0.20) is still less than MMR with the logit predictor and x-specific thresholds (0.23), and it equals MMR for both the marginal probability and the linear predictor. As shown in Section 3.1.1, MMR is zero with the correctly specified predictor and optimal thresholds $t_x = 1-U_{xB}$. This is reported in the final column of the left panel of Table SA3, demonstrating the full benefit of relaxing the constraint that thresholds be x-invariant.

*Finite-Sample MR*

Finite-sample MMR findings are reported on the right side of Table SA3. The patterns of variation across rows or across columns are qualitatively similar to the patterns for limit MR, both for x-invariant and x-specific welfare. The notable quantitative differences that arise concern the previously discussed decrease in MR of misspecified models with finite-sample data. With a correctly specified model and optimal limit thresholds, limit MMR is zero and finite-sample MMR is 0.04 in each specification of welfare. These results reinforce our findings that the analytical results for limit MR are instructive regarding finite-sample MR, and that sampling variation can be beneficial when the predictor is misspecified or suboptimal thresholds are used.

C.3. Specifications with State Space Restricted to Increasing Functions of x

Here we suppose that the conditional event probability is known to be a weakly increasing function of x. Tables SA4 to SA6 show how this restriction on the state space decreases MR for different prediction-threshold pairs.

*Limit MR*

Table SA4 shows that, with x-invariant welfare and thresholds, the state-space restriction generates notable decreases in MR for the predictors that are monotone functions of x, whereas MR of the marginal and cell probability predictors remain unchanged relative to MR with the unrestricted state space. Linear and logit prediction behave similarly in each case, yielding minimum MR of just 0.05 or 0.08. This contrasts sharply with the curious findings of very large MR for logit with the unrestricted state space and $U_{xB}$=0.2. The binary classifier assigns treatment optimally when $U_{xB}$ = 0.5, whereas MR=0.3 when $U_{xB}$=0.2, as is the case for the unrestricted state space.

Table SA6 shows that in both specifications with x-invariant welfare, $t_x = 1-U_{xB}$ minimizes MR among all x-invariant threshold vectors. For only one predictor, the marginal probability, does an x-specific threshold vector further reduce MR. With this predictor, MR falls from 0.25 (0.15) to 0.08 (0.09) with $U_{xB}$=0.5(0.2) and when $t_x$ is allowed to vary with x. For example, with $U_{xB}$=0.5, MR is minimized with the threshold vector $t_x$ = [0.8, 0.6, 0.4, 0.2].

Table SA6 summarizes and Table SA5 details findings with x-specific welfare. Linear and logit prediction again behave similarly, with the latter somewhat outperforming the former: MR with logit and $t_x = 1-U_{xB}$ is 0.04, whereas MR with linear prediction is at least twice the size. The binary classifier performs relatively well in the two welfare specifications that are symmetric about 0.5, with MR = 0.10 in one case and 0.15 in the other. The use of x-specific thresholds yields striking reductions in MR for the marginal probability predictor, falling to 0.08 from 0.35 or 0.40. In each case, MMR obtains with $t_x$ = [0.9, 0.7, 0.3, 0.1].

*Finite-Sample MR*

Finite-sample MMR findings are summarized on the right side of Table SA6. The patterns of variation are qualitatively similar to the patterns for limit MR. As we found with the unrestricted state space, finite-sample MR is often smaller than limit MR when misspecified predictors or suboptimal thresholds are used.

Supplemental Appendix: References

**Table SA1. x-Invariant Welfare $U_{xB}$ and x-Invariant Thresholds $t_x$, with the Unrestricted State Space**

| Welfare | Singleton Minimax Regret with No Data | x-Invariant Threshold t | Limit Maximum Regret | | | | | Finite-Sample Maximum Regret<br>*N = 40*, 50 Monte Carlo simulations per specification | | | | |
|---|---|---|---|---|---|---|---|---|---|---|---|---|
| | | | Marginal Probability | Linear Prediction Square Loss | Logit Prediction Square Loss | Minimum MCR | Cell Probability | Marginal Proportion | Linear Prediction Square Loss | Logit Prediction Square Loss | Minimum MCR | Cell Proportion |
| | | **0.1** | 0.39 | 0.39 | 0.39 | 0.13 | 0.35 | 0.38 | 0.33 | 0.38 | 0.14 | 0.22 |
| | | **0.2** | 0.38 | 0.34 | 0.35 | 0.13 | 0.25 | 0.38 | 0.30 | 0.32 | 0.14 | 0.14 |
| | | **0.3** | 0.31 | 0.30 | 0.31 | 0.13 | 0.15 | 0.27 | 0.27 | 0.25 | 0.14 | 0.09 |
| **$U_{xB}$ = 0.5** | | **0.4** | 0.25 | 0.34 | 0.34 | 0.13 | 0.05 | 0.25 | 0.29 | 0.25 | 0.14 | 0.05 |
| **for all x** | 0.50 | **0.5** | 0.25 | 0.34 | 0.31 | 0.13 | 0.00 | 0.25 | 0.25 | 0.25 | 0.14 | 0.04 |
| | | **0.6** | 0.25 | 0.34 | 0.26 | 0.13 | 0.10 | 0.25 | 0.28 | 0.25 | 0.14 | 0.08 |
| | | **0.7** | 0.33 | 0.30 | 0.30 | 0.13 | 0.20 | 0.30 | 0.28 | 0.26 | 0.14 | 0.13 |
| | | **0.8** | 0.38 | 0.34 | 0.35 | 0.13 | 0.30 | 0.38 | 0.31 | 0.32 | 0.14 | 0.22 |
| | | **0.9** | 0.40 | 0.40 | 0.39 | 0.13 | 0.40 | 0.38 | 0.38 | 0.38 | 0.14 | 0.29 |
| | | **0.1** | 0.69 | 0.69 | 0.69 | 0.30 | 0.65 | 0.64 | 0.59 | 0.61 | 0.25 | 0.45 |
| | | **0.2** | 0.60 | 0.60 | 0.59 | 0.30 | 0.55 | 0.60 | 0.51 | 0.55 | 0.25 | 0.36 |
| | | **0.3** | 0.54 | 0.53 | 0.53 | 0.30 | 0.45 | 0.49 | 0.43 | 0.46 | 0.25 | 0.27 |
| **$U_{xB}$ = 0.2** | | **0.4** | 0.44 | 0.44 | 0.44 | 0.30 | 0.35 | 0.40 | 0.40 | 0.40 | 0.25 | 0.21 |
| **for all x** | 0.20 | **0.5** | 0.39 | 0.41 | 0.40 | 0.30 | 0.25 | 0.34 | 0.34 | 0.40 | 0.25 | 0.15 |
| | | **0.6** | 0.29 | 0.34 | 0.40 | 0.30 | 0.15 | 0.24 | 0.26 | 0.40 | 0.25 | 0.09 |
| | | **0.7** | 0.20 | 0.24 | 0.40 | 0.30 | 0.05 | 0.20 | 0.19 | 0.40 | 0.25 | 0.05 |
| | | **0.8** | 0.15 | 0.19 | 0.40 | 0.30 | 0.00 | 0.15 | 0.16 | 0.40 | 0.25 | 0.04 |
| | | **0.9** | 0.15 | 0.15 | 0.43 | 0.30 | 0.10 | 0.15 | 0.15 | 0.40 | 0.25 | 0.07 |

Notes: The scalar x takes on four values. The distribution is uniform across those four values both for the limit and the finite-sample computations. The finite-sample Monte Carlo simulations of size N = 40 draw 10 observations at each of the four values of x, with 50 simulations performed for each specification, prediction method, threshold value $t_x$, and state $p_{sx}$ to approximate the error probabilities in equation (16). For each prediction method but Minimium Misclassification Rate (MCR), shaded entries denote the constrained minimum of computed MR across the x-invariant threshold values. Entries for the binary {0, 1} MCR classifier are invariant across thresholds. No such entries are shaded.

**Table SA2. x-Specific Welfare $U_{xB}$ and x-Invariant Thresholds $t_x$, with the Unrestricted State Space**

| Welfare | Singleton Minimax Regret with No Data | x-Invariant Threshold t | Limit Maximum Regret | | | | | Finite-Sample Maximum Regret<br>*N = 40*, 50 Monte Carlo simulations per specification | | | | |
|---|---|---|---|---|---|---|---|---|---|---|---|---|
| | | | Marginal Probability | Linear Prediction Square Loss | Logit Prediction Square Loss | Minimum MCR | Cell Probability | Marginal Proportion | Linear Prediction Square Loss | Logit Prediction Square Loss | Minimum MCR | Cell Proportion |
| | | **0.1** | 0.45 | 0.39 | 0.40 | 0.26 | 0.35 | 0.45 | 0.34 | 0.40 | 0.22 | 0.23 |
| | | **0.2** | 0.45 | 0.39 | 0.38 | 0.26 | 0.26 | 0.45 | 0.32 | 0.34 | 0.22 | 0.18 |
| | | **0.3** | 0.39 | 0.39 | 0.39 | 0.26 | 0.21 | 0.35 | 0.33 | 0.27 | 0.22 | 0.14 |
| | | **0.4** | 0.35 | 0.41 | 0.40 | 0.26 | 0.18 | 0.35 | 0.32 | 0.29 | 0.22 | 0.12 |
| **$U_{xB}$ = [0.2, 0.4, 0.6, 0.8]** | 0.30 | **0.5** | 0.35 | 0.41 | 0.36 | 0.26 | 0.18 | 0.35 | 0.30 | 0.30 | 0.22 | 0.12 |
| | | **0.6** | 0.35 | 0.41 | 0.38 | 0.26 | 0.19 | 0.35 | 0.32 | 0.30 | 0.22 | 0.15 |
| | | **0.7** | 0.40 | 0.39 | 0.38 | 0.26 | 0.24 | 0.37 | 0.32 | 0.30 | 0.22 | 0.18 |
| | | **0.8** | 0.45 | 0.39 | 0.38 | 0.26 | 0.30 | 0.45 | 0.32 | 0.34 | 0.22 | 0.23 |
| | | **0.9** | 0.45 | 0.40 | 0.40 | 0.26 | 0.40 | 0.45 | 0.40 | 0.40 | 0.22 | 0.32 |
| | | **0.1** | 0.48 | 0.44 | 0.44 | 0.31 | 0.36 | 0.48 | 0.38 | 0.43 | 0.24 | 0.27 |
| | | **0.2** | 0.48 | 0.44 | 0.43 | 0.31 | 0.31 | 0.48 | 0.35 | 0.35 | 0.24 | 0.22 |
| | | **0.3** | 0.41 | 0.45 | 0.45 | 0.31 | 0.28 | 0.40 | 0.35 | 0.31 | 0.24 | 0.19 |
| | | **0.4** | 0.40 | 0.45 | 0.45 | 0.31 | 0.28 | 0.40 | 0.38 | 0.38 | 0.24 | 0.18 |
| **$U_{xB}$ = [0.1, 0.3, 0.7, 0.9]** | 0.20 | **0.5** | 0.40 | 0.45 | 0.45 | 0.31 | 0.28 | 0.40 | 0.35 | 0.35 | 0.24 | 0.18 |
| | | **0.6** | 0.40 | 0.45 | 0.45 | 0.31 | 0.28 | 0.40 | 0.39 | 0.38 | 0.24 | 0.19 |
| | | **0.7** | 0.43 | 0.45 | 0.45 | 0.31 | 0.29 | 0.40 | 0.34 | 0.32 | 0.24 | 0.22 |
| | | **0.8** | 0.48 | 0.44 | 0.43 | 0.31 | 0.34 | 0.48 | 0.34 | 0.36 | 0.24 | 0.26 |
| | | **0.9** | 0.48 | 0.44 | 0.43 | 0.31 | 0.40 | 0.48 | 0.43 | 0.43 | 0.24 | 0.33 |

Notes: The scalar x takes on four values. The distribution is uniform across those four values both for the limit and the finite-sample computations. The finite-sample Monte Carlo simulations of size N = 40 draw 10 observations at each of the four values of x, with 50 simulations performed for each specification, prediction method, threshold value $t_x$, and state $p_{sx}$ to approximate the error probabilities in equation (16). For each prediction method but Minimium Misclassification Rate (MCR), shaded entries denote the constrained minimum of computed MR across the x-invariant threshold values. Entries for the binary {0, 1} MCR classifier are invariant across thresholds. No such entries are shaded.

**Table SA3. Constrained Minimax Regret with x-Invariant or x-Specific Thresholds, with the Unrestricted State Space**

| Welfare | Singleton Minimax Regret with No Data | Threshold | Limit Maximum Regret | | | | | Finite-Sample Maximum Regret<br>$N = 40$, 50 simulations per specification | | | | |
|---|---|---|---|---|---|---|---|---|---|---|---|---|
| | | | Marginal Probability | Linear Prediction Square Loss | Logit Prediction Square Loss | Minimum MCR | Cell Probability | Marginal Proportion | Linear Prediction Square Loss | Logit Prediction Square Loss | Minimum MCR | Cell Proportion |
| **$U_{xB}$ = 0.5 for all x** | 0.50 | x-Invariant t MMR | 0.25 | 0.30 | 0.26 | 0.13 | 0.00 | 0.25 | 0.25 | 0.25 | 0.14 | 0.04 |
| | | $t_x = 1 - U_{xB}$ | 0.25 | 0.34 | 0.31 | | 0.00 | 0.25 | 0.25 | 0.25 | | 0.04 |
| | | x-Specific $t_x$ MMR | 0.25 | 0.29 | 0.26 | | 0.00 | 0.25 | 0.25 | 0.25 | | 0.04 |
| **$U_{xB}$ = 0.2 for all x** | 0.20 | x-Invariant t MMR | 0.15 | 0.15 | 0.40 | 0.30 | 0.00 | 0.15 | 0.15 | 0.40 | 0.25 | 0.04 |
| | | $t_x = 1 - U_{xB}$ | 0.15 | 0.19 | 0.40 | | 0.00 | 0.15 | 0.16 | 0.40 | | 0.04 |
| | | x-Specific $t_x$ MMR | 0.15 | 0.15 | 0.40 | | 0.00 | 0.15 | 0.15 | 0.40 | | 0.03 |
| **$U_{xB}$ = [0.2, 0.4, 0.6, 0.8]** | 0.30 | x-Invariant t MMR | 0.35 | 0.39 | 0.36 | 0.26 | 0.18 | 0.35 | 0.30 | 0.30 | 0.22 | 0.12 |
| | | $t_x = 1 - U_{xB}$ | 0.30 | 0.28 | 0.35 | | 0.00 | 0.30 | 0.22 | 0.22 | | 0.04 |
| | | x-Specific $t_x$ MMR | 0.30 | 0.21 | 0.23 | | 0.00 | 0.30 | 0.20 | 0.20 | | 0.04 |
| **$U_{xB}$ = [0.1, 0.3, 0.7, 0.9]** | 0.20 | x-Invariant t MMR | 0.40 | 0.44 | 0.43 | 0.31 | 0.28 | 0.40 | 0.34 | 0.32 | 0.24 | 0.18 |
| | | $t_x = 1 - U_{xB}$ | 0.20 | 0.20 | 0.23 | | 0.00 | 0.20 | 0.20 | 0.20 | | 0.04 |
| | | x-Specific $t_x$ MMR | 0.20 | 0.20 | 0.23 | | 0.00 | 0.20 | 0.19 | 0.17 | | 0.04 |

Notes: Entries report MMR for each prediction method constraining the thresholds to be (i) x-invariant with t taking on any one of the nine values reported in Tables SA1 and SA2, (ii) the optimal threshold vector when the cell probability $p_{sx}$ is known ($t_x = 1 - U_{xB}$), or (iii) x-specific $t_x$ with the vector of thresholds taking on any of the $9^4$ = 6,561 combinations of those same nine values in Tables SA1 and SA2.

**Table SA4. x-Invariant Welfare $U_{xB}$ and x-Invariant Thresholds, with the Restricted State Space**

| Welfare | Singleton Minimax Regret with No Data | x-Invariant Threshold t | Limit Maximum Regret | | | | | Finite-Sample Maximum Regret *N = 40*, 50 Monte Carlo simulations per specification | | | | |
|---|---|---|---|---|---|---|---|---|---|---|---|---|
| | | | Marginal Probability | Linear Prediction Square Loss | Logit Prediction Square Loss | Minimum MCR | Cell Probability | Marginal Proportion | Linear Prediction Square Loss | Logit Prediction Square Loss | Minimum MCR | Cell Proportion |
| | | **0.1** | 0.39 | 0.36 | 0.36 | 0.00 | 0.35 | 0.38 | 0.27 | 0.27 | 0.03 | 0.21 |
| | | **0.2** | 0.38 | 0.26 | 0.26 | 0.00 | 0.25 | 0.38 | 0.19 | 0.19 | 0.03 | 0.15 |
| | | **0.3** | 0.31 | 0.16 | 0.16 | 0.00 | 0.15 | 0.26 | 0.13 | 0.12 | 0.03 | 0.09 |
| **$U_{xB}$ = 0.5** | | **0.4** | 0.25 | 0.11 | 0.08 | 0.00 | 0.05 | 0.25 | 0.10 | 0.06 | 0.03 | 0.05 |
| **for all x** | 0.50 | **0.5** | 0.25 | 0.08 | 0.05 | 0.00 | 0.00 | 0.25 | 0.06 | 0.04 | 0.03 | 0.04 |
| | | **0.6** | 0.25 | 0.13 | 0.08 | 0.00 | 0.10 | 0.25 | 0.13 | 0.06 | 0.03 | 0.08 |
| | | **0.7** | 0.33 | 0.20 | 0.20 | 0.00 | 0.20 | 0.28 | 0.13 | 0.11 | 0.03 | 0.14 |
| | | **0.8** | 0.38 | 0.30 | 0.26 | 0.00 | 0.30 | 0.38 | 0.20 | 0.18 | 0.03 | 0.21 |
| | | **0.9** | 0.40 | 0.40 | 0.36 | 0.00 | 0.40 | 0.38 | 0.27 | 0.27 | 0.03 | 0.29 |
| | | **0.1** | 0.69 | 0.66 | 0.66 | 0.30 | 0.65 | 0.61 | 0.54 | 0.54 | 0.19 | 0.45 |
| | | **0.2** | 0.60 | 0.56 | 0.56 | 0.30 | 0.55 | 0.60 | 0.44 | 0.44 | 0.19 | 0.36 |
| | | **0.3** | 0.54 | 0.46 | 0.46 | 0.30 | 0.45 | 0.47 | 0.35 | 0.35 | 0.19 | 0.28 |
| **$U_{xB}$ = 0.2** | | **0.4** | 0.44 | 0.36 | 0.36 | 0.30 | 0.35 | 0.40 | 0.26 | 0.26 | 0.19 | 0.20 |
| **for all x** | 0.20 | **0.5** | 0.39 | 0.26 | 0.26 | 0.30 | 0.25 | 0.33 | 0.18 | 0.19 | 0.19 | 0.14 |
| | | **0.6** | 0.29 | 0.16 | 0.16 | 0.30 | 0.15 | 0.23 | 0.12 | 0.12 | 0.19 | 0.08 |
| | | **0.7** | 0.20 | 0.06 | 0.06 | 0.30 | 0.05 | 0.20 | 0.07 | 0.07 | 0.19 | 0.05 |
| | | **0.8** | 0.15 | 0.05 | 0.05 | 0.30 | 0.00 | 0.15 | 0.05 | 0.05 | 0.19 | 0.03 |
| | | **0.9** | 0.15 | 0.10 | 0.08 | 0.30 | 0.10 | 0.15 | 0.10 | 0.05 | 0.19 | 0.07 |

Notes: The scalar x takes on four values. The distribution is uniform across those four values both for the limit and the finite-sample computations. The finite-sample Monte Carlo simulations of size N = 40 draw 10 observations at each of the four values of x, with 50 simulations performed for each specification, prediction method, threshold value $t_x$, and state $p_{sx}$ to approximate the error probabilities in equation (16). For each prediction method but Minimium Misclassification Rate (MCR), shaded entries denote the constrained minimum of computed MR across the x-invariant threshold values. Entries for the binary {0, 1} MCR classifier are invariant across thresholds. No such entries are shaded.

**Table SA5. x-Specific Welfare $U_{xB}$ and x-Invariant Thresholds $t_x$, with the Restricted State Space**

| Welfare | Singleton Minimax Regret with No Data | x-Invariant Threshold t | Limit Maximum Regret | | | | | Finite-Sample Maximum Regret<br>*N = 40*, 50 Monte Carlo simulations per specification | | | | |
|---|---|---|---|---|---|---|---|---|---|---|---|---|
| | | | Marginal Probability | Linear Prediction Square Loss | Logit Prediction Square Loss | Minimum MCR | Cell Probability | Marginal Proportion | Linear Prediction Square Loss | Logit Prediction Square Loss | Minimum MCR | Cell Proportion |
| **$U_{xB}$ = [0.2, 0.4, 0.6, 0.8]** | 0.30 | **0.1** | 0.45 | 0.36 | 0.36 | 0.10 | 0.35 | 0.45 | 0.27 | 0.27 | 0.10 | 0.22 |
| | | **0.2** | 0.45 | 0.28 | 0.28 | 0.10 | 0.26 | 0.45 | 0.21 | 0.20 | 0.10 | 0.16 |
| | | **0.3** | 0.39 | 0.20 | 0.20 | 0.10 | 0.19 | 0.35 | 0.16 | 0.16 | 0.10 | 0.12 |
| | | **0.4** | 0.35 | 0.16 | 0.14 | 0.10 | 0.13 | 0.35 | 0.12 | 0.12 | 0.10 | 0.11 |
| | | **0.5** | 0.35 | 0.13 | 0.09 | 0.10 | 0.10 | 0.35 | 0.11 | 0.11 | 0.10 | 0.11 |
| | | **0.6** | 0.35 | 0.16 | 0.14 | 0.10 | 0.15 | 0.35 | 0.11 | 0.11 | 0.10 | 0.13 |
| | | **0.7** | 0.40 | 0.23 | 0.23 | 0.10 | 0.23 | 0.36 | 0.20 | 0.15 | 0.10 | 0.16 |
| | | **0.8** | 0.45 | 0.30 | 0.30 | 0.10 | 0.30 | 0.45 | 0.21 | 0.20 | 0.10 | 0.23 |
| | | **0.9** | 0.45 | 0.40 | 0.38 | 0.10 | 0.40 | 0.45 | 0.28 | 0.28 | 0.10 | 0.29 |
| **$U_{xB}$ = [0.1, 0.3, 0.7, 0.9]** | 0.20 | **0.1** | 0.48 | 0.38 | 0.38 | 0.15 | 0.36 | 0.48 | 0.29 | 0.29 | 0.15 | 0.24 |
| | | **0.2** | 0.48 | 0.30 | 0.30 | 0.15 | 0.29 | 0.48 | 0.24 | 0.24 | 0.15 | 0.19 |
| | | **0.3** | 0.41 | 0.24 | 0.24 | 0.15 | 0.23 | 0.40 | 0.19 | 0.19 | 0.15 | 0.19 |
| | | **0.4** | 0.40 | 0.19 | 0.19 | 0.15 | 0.18 | 0.40 | 0.18 | 0.18 | 0.15 | 0.17 |
| | | **0.5** | 0.40 | 0.15 | 0.14 | 0.15 | 0.15 | 0.40 | 0.16 | 0.15 | 0.15 | 0.16 |
| | | **0.6** | 0.40 | 0.20 | 0.19 | 0.15 | 0.20 | 0.40 | 0.17 | 0.17 | 0.15 | 0.18 |
| | | **0.7** | 0.43 | 0.25 | 0.25 | 0.15 | 0.25 | 0.40 | 0.23 | 0.21 | 0.15 | 0.21 |
| | | **0.8** | 0.48 | 0.33 | 0.35 | 0.15 | 0.33 | 0.48 | 0.25 | 0.24 | 0.15 | 0.25 |
| | | **0.9** | 0.48 | 0.40 | 0.40 | 0.15 | 0.40 | 0.48 | 0.30 | 0.30 | 0.15 | 0.31 |

Notes: The scalar x takes on four values. The distribution is uniform across those four values both for the limit and the finite-sample computations. The finite-sample Monte Carlo simulations of size N = 40 draw 10 observations at each of the four values of x, with 50 simulations performed for each specification, prediction method, threshold value $t_x$, and state $p_{sx}$ to approximate the error probabilities in equation (16). For each prediction method but Minimium Misclassification Rate (MCR), shaded entries denote the constrained minimum of computed MR across the x-invariant threshold values. Entries for the binary {0, 1} MCR classifier are invariant across thresholds. No such entries are shaded.

**Table SA6. Constrained Minimax Regret with x-Invariant or x-Specific Thresholds, with the Restricted State Space**

| Welfare | Singleton Minimax Regret with No Data | Threshold | Limit Maximum Regret | | | | | Finite-Sample Maximum Regret<br>*N=40*, 50 simulations per specification | | | | |
|---|---|---|---|---|---|---|---|---|---|---|---|---|
| | | | Marginal Probability | Linear Prediction Square Loss | Logit Prediction Square Loss | Minimum MCR | Cell Probability | Marginal Proportion | Linear Prediction Square Loss | Logit Prediction Square Loss | Minimum MCR | Cell Proportion |
| **$U_{xB}$ = 0.5 for all x** | 0.50 | x-Invariant t MMR | 0.25 | 0.08 | 0.05 | 0.00 | 0.00 | 0.25 | 0.06 | 0.04 | 0.03 | 0.04 |
| | | $t_x = 1 - U_{xB}$ | 0.25 | 0.08 | 0.05 | | 0.00 | 0.25 | 0.06 | 0.04 | | 0.04 |
| | | x-Specific $t_x$ MMR | 0.08 | 0.08 | 0.05 | | 0.00 | 0.06 | 0.06 | 0.04 | | 0.03 |
| **$U_{xB}$ = 0.2 for all x** | 0.20 | x-Invariant t MMR | 0.15 | 0.05 | 0.05 | 0.30 | 0.00 | 0.15 | 0.05 | 0.05 | 0.19 | 0.03 |
| | | $t_x = 1 - U_{xB}$ | 0.15 | 0.05 | 0.05 | | 0.00 | 0.15 | 0.05 | 0.05 | | 0.03 |
| | | x-Specific $t_x$ MMR | 0.09 | 0.05 | 0.05 | | 0.00 | 0.07 | 0.05 | 0.04 | | 0.03 |
| **$U_{xB}$ = [0.2, 0.4, 0.6, 0.8]** | 0.30 | x-Invariant t MMR | 0.35 | 0.13 | 0.09 | 0.10 | 0.10 | 0.35 | 0.11 | 0.11 | 0.10 | 0.11 |
| | | $t_x = 1 - U_{xB}$ | 0.15 | 0.10 | 0.04 | | 0.00 | 0.11 | 0.10 | 0.03 | | 0.03 |
| | | x-Specific $t_x$ MMR | 0.08 | 0.08 | 0.04 | | 0.00 | 0.06 | 0.05 | 0.03 | | 0.02 |
| **$U_{xB}$ = [0.1, 0.3, 0.7, 0.9]** | 0.20 | x-Invariant t MMR | 0.40 | 0.15 | 0.14 | 0.15 | 0.15 | 0.40 | 0.16 | 0.15 | 0.15 | 0.16 |
| | | $t_x = 1 - U_{xB}$ | 0.08 | 0.08 | 0.04 | | 0.00 | 0.05 | 0.08 | 0.03 | | 0.02 |
| | | x-Specific $t_x$ MMR | 0.08 | 0.08 | 0.04 | | 0.00 | 0.05 | 0.06 | 0.03 | | 0.02 |

Notes: Entries report MMR for each prediction method constraining the thresholds to be (i) x-invariant with t taking on any one of the nine values reported in Tables SA4 and SA5, (ii) the optimal threshold vector when the cell probability $p_{sx}$ is known ($t_x = 1 - U_{xB}$), or (iii) x-specific $t_x$ with the vector of thresholds taking on any of the $9^4$ = 6,561 combinations of those same nine values in Tables SA4 and SA5.